# Unprecedentedly Wide Curie-Temperature Windows as Phase-Transition Design Platform for Tunable Magneto-Multifunctional Materials

*Zhi-Yang Wei, En-Ke Liu,* Yong Li, Gui-Zhou Xu, Xiao-Ming Zhang, Guo-Dong Liu, Xue-Kui Xi, Hong-Wei Zhang, Wen-Hong Wang,* Guang-Heng Wu, and Xi-Xiang Zhang*

Z.-Y. Wei, Dr. E.-K. Liu, Y. Li, G.-Z. Xu, X.-M. Zhang, Dr. H.-W. Zhang, Dr. X.-K. Xi, Dr. W.-H. Wang, Prof. G.-H. Wu

State Key Laboratory for Magnetism

Beijing National Laboratory for Condensed Matter Physics

Institute of Physics

Chinese Academy of Sciences

Beijing 100190 , P.R. China

E-mail: ekliu@iphy.ac.cn; wenhong.wang@iphy.ac.cn

Y. Li, Prof. G.-D. Liu

School of Materials Science and Engineering

Hebei University of Technology

Tianjin 300130, China

Prof. X.-X. Zhang

Division of Physical Science and Engineering

King Abdullah University of Science and Technology

Thuwal 23955-6900, Saudi Arabia.

Abstract

Employing the principle of isostructural alloying, a series of unprecedentedly wide Curie-temperature windows (CTWs), covering the ultimate temperature range of first-order magnetostructural phase transitions (MSTs), are realized between 40 and 450 K for the strongly-coupled MSTs in a single host system $Mn_{1-y}Fe_yNiGe_{1-x}Si_x$. Throughout the wide CTWs, the highly tunable MSTs show large magnetization jump, low-field effects, high-temperature giant magnetocaloric effects and robust functional stability, providing important properties to phase-transition-based magneto-multifunctional applications, including field-driven shape memory, magnetic cooling/heating and energy conversion. The unprecedentedly wide CTWs open up a design platform for magneto-multifunctional multiferroic alloys that can be manipulated in a quite large temperature space in various scales and patterns, and by multiple physical fields.

A coupling of ferroelasticity and ferromagnetism[1, 2] can lead to a multiferroic behavior of first-order magnetostructural phase transition (MST) in magnetoelastic materials. Attractive physical effects, such as ferromagnetic shape memory,[3] magneto-strain,[4] magnetocaloric effect (MCE),[5, 6] magnetoresistance,[7] and exchange bias,[8] are observed based on the MSTs. These effects are receiving increasing attentions from the applications in actuating,[9] sensing,[10] magnetic cooling,[11] heat pump[12] and energy conversion.[13] As an important class of MSTs, ferromagnetic martensitic transformations (FMMTs) are widely found in Heusler, Fe-based, and MM'X alloys, and produce diversiform physical discontinuities due to the significant alterations in crystallographic, electronic, orbital and magnetic structures in the systems. Extraordinary magneto-multifunctional properties emerge thereby and can be tuned in different ways.

Profiting from FMMTs, the magnetic-field-induced shape memory effect and giant magneto-strain/magnetostriction have been extensively studied with promising potential in micro-mechanical controls and strain outputs. The magnetocaloric effect (MCE),[14] which happens in a magnetic transition, can be enhanced appreciably by first-order FMMTs with great changes in structural entropy in spin-lattice coupled systems.[15-17] Materials bearing FMMTs are thus further considered as candidates for caloric applications, probably combined with mechanocaloric or electrocaloric effects.[18, 19] Very recently, an inspiring application of electric power generation,[20, 21] using first-order FMMT ferromagnets driven by high-temperature heat resources, becomes increasingly attractive for the energy conversion, which demonstrates an exciting advance in magneto-multifunctionalities of magnetoelastic alloys.

For all the magneto-multifunctionalities, the magnetostructural coupling plays a fundamental role since only in this case can the structural and magnetic transitions modulate each other. In order to approach FMMTs, the martensitic transformations should happen below the Curie temperature ($T_C$) of the system so that the transition couples with magnetic state changes. Along with the increasing demands of diversified functional applications, searching for new materials with highly tunable FMMTs, especially in a much wider working temperature range, is ever growing. However, challenges exist in practice. One challenge is to enlarge the magnetization change ($\Delta$M) across FMMTs to maximize the magnetic-energy change especially in a moderate magnetic field. The other challenge is to broaden the temperature range where the FMMTs can take place, similar to the case in magnetocaloric materials Gd-Si-Ge between 20 and 290 K,[23] or the case in Mn-Fe-P-As between 150 and 350 K.[24] For first-order FMMTs, the structural transitions are often limited in a temperature range between just above room and liquid-nitrogen temperatures. If $T_C$ of an FMMT material could be tailored to higher temperatures, the temperature space in which magnetostructural coupling can occur will be expanded largely.

As a rising family of FMMT materials, the hexagonal MM'X (M, M' = transition metals, X = carbon or boron group elements) compounds have been intensively studied in recent years.[25-27] The FMMTs in these compounds are characterized by small thermal hysteresis, high temperature-sensitivity, high Curie-temperatures, gigantic anisotropic strains, huge volume expansions, large magnetization jump, magnetic-field-induced shape memory effects, and giant MCEs,[26, 28-30] which collectively make MM'X compounds rather different from the conventional FMMT materials (for example, Heusler alloys). Many methods, including physical pressure, chemical modification, atom-vacancy introducing, and quench-relaxation annealing, have been adopted to tune the FMMTs. In our previous works,[26, 29] we proposed an isostructural alloying principle to guide the chemical substitution during the material design. On this principle, we alloyed two isostructural compounds that have the same crystal structure but distinct phase stability and different magnetic behaviors together to form a new compound, by which we were able to manipulate both the phase transition and the magnetic exchange interactions simultaneously in a material host. Recently, this effective principle has generated active influences on design and realization of the desired MST materials.[30-36]

In the light of this principle, in our previous work[26] we have alloyed MnNiGe with isostructural FeNiGe ($Mn_{1-y}Fe_yNiGe$) and established a stable magnetostructural coupling in a broad Curie-temperature window (CTW), as illustrated in left diagram in **Scheme 1**, in which tunable large MCEs have been obtained. Upon increasing Fe content ($y$), $T_t$ of FMMT was efficiently lowered below $T_C^M$ of martensite (point B in **Scheme 1**) and was continuously decreased down to the magnetic-frozen temperature ($T_g^A$, point C in **Scheme 1**) of the spin-glass-like austenite. Simultaneously, for martensite phase introducing of Fe atom gradually converted the spirally antiferromagnetic (AFM) coupling to ferromagnetic (FM) one, with $T_N^M$ becoming $T_C^M$, and finally resulted in a high magnetization when $y > 0.22$ (near point C at $y_1 = 0.26$, deep blue region in the left diagram in **Scheme 1**). As a result, a CTW was established between $T_C^M \sim 350$ K and $T_g^A \sim 70$ K. In the CTW, the strong ferromagnetism of martensite, featured by *decreasing* saturation field ($H_S$) and *increasing* saturation magnetization ($M_S$), shows a positive correlation with Fe content. This work[26] provides a fundamental understanding on the tuning of magnetostructural coupling under the isostructural alloying principle and opens up a way to explore large CTWs for strongly-coupled MST materials.

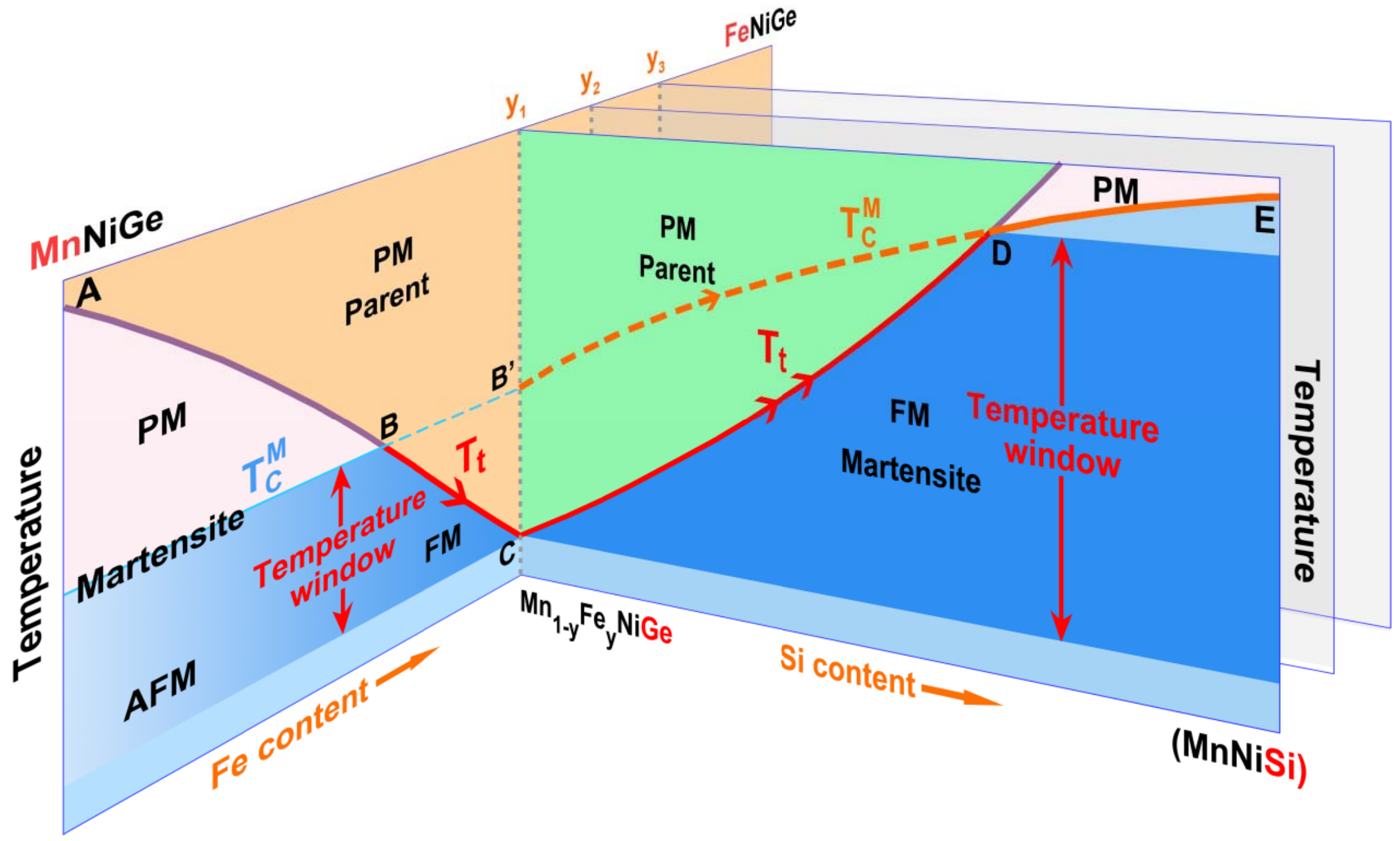

**Scheme 1.** A schematic of the alloy design indicated by the structural and magnetic phase diagram for MnNiGe-FeNiGe ($Mn_{1-y}Fe_yNiGe$) isostructural system (Left) and $Mn_{1-y}Fe_yNiGe$-MnNiSi ($Mn_{1-y}Fe_yNiGe_{1-x}Si_x$ termed as ($y$, $x$)) one (Right). Curve ABC represents $T_t$ curve of martensitic structural transition from $Ni_2In$-type hexagonal parent phase to TiNiSi-type orthorhombic martensite. $T_t$ curve crosses Curie-temperature curve ($T_C^M$) of martensite at point B and magnetic ordering temperature ($T_g^A$) of austenite at point C. Between $T_C^M$ and $T_g^A$ (points B and C), a Curie-temperature window (CTW) and tunable PM-FM-type FMMTs were realized by Liu *et al.*[26] Right part of the schematic shows the desired phase diagram in this study produced by isostructurally alloying $Mn_{1-y}Fe_yNiGe$ with MnNiSi. In this phase diagram, $T_t$ and $T_C^M$ are simultaneously expected to increase and cross over (point D) at high temperatures. A much wider CTW (between points C and D) is highly desired. Various Fe content ($y_1$, $y_2$,...) corresponds to a series of expected phase diagrams and CTWs.

In this work, in line with the important success mentioned above, we design new material systems step by step by taking the strategy shown in **Scheme 1**. If one keeps the strong ferromagnetism around point C, at the same time tailors the transition ($T_t$) to high temperatures, one may obtain a series of strongly-coupled MSTs with low-field effects and large ΔM. $T_C^M$, which determines the upper critical temperature ($T_{cr}^{up}$, point D) of the CTW, should be simultaneously raised from point B' to high temperatures. By achieving this, abundant magneto-multifunctional applications can be expanded and promoted freely over a very broad temperature range from cryogenic to high temperatures.

According to the isostructural alloying principle, to realize the above purpose the isostructural counterpart to $Mn_{1-y}Fe_yNiGe$ system must meet three conditions: 1) A high $T_t$, which allows us to tailor the FMMT ($T_t$) of the alloyed system from low to high temperatures; 2) A high $T_C^M$, to upraise the $T_{cr}^{up}$ of CTW; 3) A strong ferromagnetic coupling in martensite phase, to gain large ΔM, low $H_S$ and desired magneto-multifunctional effects. Among $Ni_2In$-type hexagonal compounds, MnNiSi can be identified as a promising isostructural counterpart. Both its $T_t$ and $T_C^M$ are as high as 1200 K and 600 K, respectively.[37, 38] Its martensite phase is a strong ferromagnet with a low $H_S$ and a large $M_S$. The high $T_t$ hopefully drives up the FMMT from point C (**Scheme 1**) continuously to high temperatures with increasing alloying level of Si content ($x$) (red line with dual-arrows, **Scheme 1**). Similarly, $T_C^M$ may be upraised as well. It is desirable that $T_t$ and $T_C^M$ curves would cross over at high temperatures, resulting in a brand-new CTW between two far-apart points C and D.

After determining the isostructural counterpart MnNiSi, $Mn_{0.74}Fe_{0.26}NiGe$ was taken as the first starting alloy. For MnNiSi, we just alloyed it with $Mn_{0.74}Fe_{0.26}NiGe$ by simply substituting Ge for Si atoms. An isostructural system of $Mn_{0.74}Fe_{0.26}NiGe_{1-x}Si_x$ was created. Although the FMMTs vanish in $Mn_{1-y}Fe_yNiGe$ compounds when $y > 0.26$ (for example, $y$ = 0.36, 0.46, 0.55), these compositions were further taken as our starting alloys, with a consideration that an intrinsic strong ferromagnetism can be expected owing to the ferromagnetic coupling between Fe-Mn atoms in the martensite form.[26] In this work, dual-variable $Mn_{1-y}Fe_yNiGe_{1-x}Si_x$ ($y$ = 0.26, 0.36, 0.46, 0.55; $0 \le x \le 1$) systems were studied systematically. A crystallographic and magnetic phase diagram was proposed in **Figure 1a** and **1b**, based on the data from the structural (XRD), magnetic (M(T) curve) and thermal (DSC) measurements (more data in Supporting Information, **Figures S1, S3, S5 and S6**).

The alloy series of ($y$ = 0.26, $x$) is taken as an example (**Figure 1a**). Upon introducing Si atoms on Ge sites, $T_t$ begins to increase from low temperature of 74 K for $x$ = 0 to high temperature up to 1000 K for $x$ = 1.0, which is very close to $T_t$ (~ 1200 K) of stoichiometric MnNiSi.[37] Meanwhile, $T_C^M$ is also increased, indicating Si substitution results in a significant enhancement in magnetic exchange interactions between Mn/Fe-Mn/Fe atoms in martensite phase. One can further see that the slope of $T_t$-(Si) $x$ curve is much larger than that of $T_C^M$-(Si) $x$ one. This implies that the Si substitution imposes more influence on the structural transition than on the magnetic coupling. The crossover point D mentioned in **Scheme 1** appears at 434 K, which is much higher than $T_C^M$ (350 K) of starting alloy $Mn_{1-y}Fe_yNiGe$.[26] For higher Fe contents ($y$ = 0.36, 0.46, 0.55), a similar behavior is observed. Both $T_t$ and $T_C^M$ are simultaneously raised to high temperatures simply by Si substitution and the crossover points are thus obtained above 400 K.

An amplified phase diagram is depicted in **Figure 1b**. For the alloy series of ($y$ = 0.26, $x$), a CTW with a width of 360 K between the lower critical temperature ($T_{cr}^{dn}$) ~ 74 K and $T_{cr}^{up}$ ~ 434 K is obtained. From **Figure 1b**, one sees that $T_{cr}^{up}$ (crossover point) goes along a kappa-like curve with a maximum value 448 K in ($y$ = 0.46, $x$) series, while with increasing Si content $T_{cr}^{dn}$ further decreases from 74 K to about 38 K. These $T_{cr}^{dn}$, below which the displacements of alloy atoms will be frozen, are the lowest transition temperatures reported in first-order FMMT systems. As an important result, a series of unprecedentedly wide CTWs with a maximal width of 405 K between $T_{cr}^{dn}$ and $T_{cr}^{up}$ are obtained in $Mn_{1-y}Fe_yNiGe_{1-x}Si_x$ system (also see inset to **Figure 1b**). These CTWs span from below liquid nitrogen temperature (~ 38 K) to room temperature and to the highest temperature of 448 K, which are much wider than the temperature distribution ranges of many other FMMTs.[17, 22-24, 27-31, 33-36] This is the first realization that highly tunable MSTs can be gained in such wide temperature range in one material system. A strong magnetostructural coupling of ferroelasticity and ferromagnetism in $Mn_{1-y}Fe_yNiGe_{1-x}Si_x$ system is thus obtained over a quite large range, connecting the ultralow and the high temperatures. In **Figure 1b**, between the window boundaries ($T_{cr}^{dn}$-$T_{cr}^{up}$) (the light-yellow region), the strongly-coupled MSTs can be obtained in any composition point with random ($y$, $x$) values.

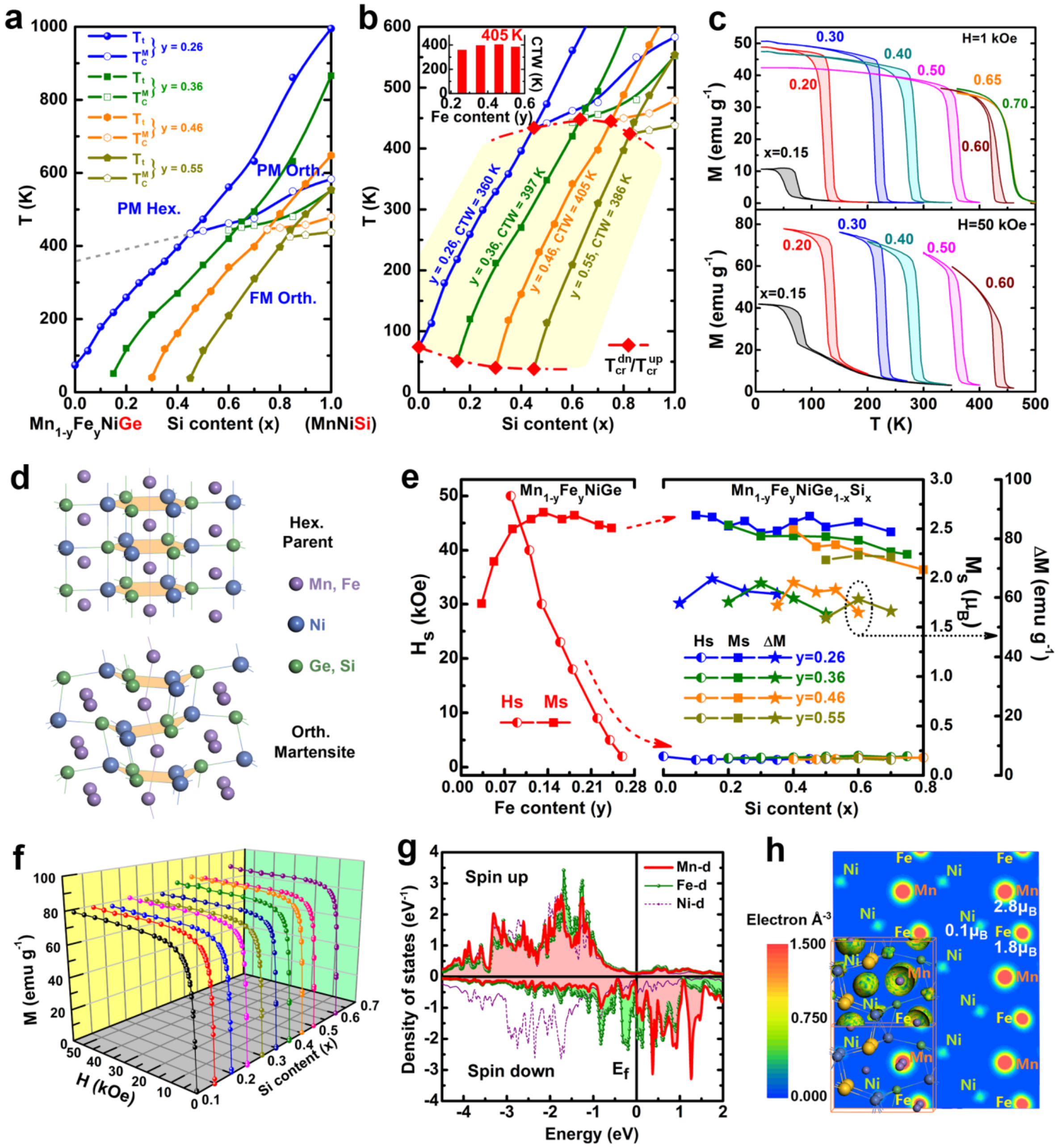

**Figure 1.** (a) Structural and magnetic phase diagram of $Mn_{1-y}Fe_yNiGe_{1-x}Si_x$ ($y$, $x$) system. (b) CTWs constructed by different Fe and Si contents. Inset shows the Fe-content ($y$) dependence of CTW width. Red diamond symbol denotes the upper and lower critical temperatures ($T_{cr}^{up}/T_{cr}^{dn}$) of CTWs. (c) Thermomagnetic curves of ($y$ = 0.36, $x$) series in magnetic field of 1 kOe (upper panel) and 50 kOe (lower panel). (d) Martensitic structural transition from hexagonal parent phase to orthorhombic martensite. (e) Saturation field and saturation magnetization of $Mn_{1-y}Fe_yNiGe$ (left panel, data from ref. [26]) and saturation field, saturation magnetization and magnetization difference (ΔM) across the FMMTs (right panel). (f) Si-content ($x$) dependence of isothermal magnetization curves at 5 K for ($y$ = 0.26, $x$) series. (g) Partial density of states (PDOS) and (h) contours of spin electron density of ($y$ = 0.5, $x$ = 0.5) alloy.

The MST behaviors within the CTWs are presented by magnetization measurements. **Figure 1c** shows the thermomagnetic M(T) curves of ($y$ = 0.36, $x$) alloy series (more data in Supporting Information, **Figure S3**). The M(T) curves with abrupt magnetization jump from paramagnetic (PM) state to FM one, and clear but small thermal hysteresis demonstrate the first-order FMMT behavior between $0.15 \le x \le 0.60$. During the Si substitution, the FMMT can be tailored from 51 K to high temperatures, showing a high tunability. These transitions involve two structures, $Ni_2In$-type hexagonal parent phase and TiNiSi-type orthorhombic martensite, as illustrated in **Figure 1d** (see structural analysis by XRD in Supporting Information, **Figure S1**). The FMMT occurs from the parent phase to the martensite via distortions of Ni-Ge/Si hexagonal rings and Mn/Fe-Mn/Fe zigzag chains, with a giant volume expansion of around 3% (more data in Supporting Information, **Figure S2 and Table S1**). Based on this structural transition, a high temperature-sensitivity and a large magnetization jump in a very narrow temperature region can be observed. Furthermore, very high magnetizations are seen even in a low field of 1 kOe. Large ΔMs across the FMMTs up to 65 emu/g are measured in the CTW, as shown in **Figure 1e** (more data in Supporting Information, **Table S2**). These ideal features will promote the magneto-multifunctional properties such as giant MCEs and electric power generation at low fields with low energy consumptions, especially in a wide temperature range including the important high-temperature region. For higher Si substitution with $x$ = 0.65 and 0.70, the thermal hysteresis becomes zero in the M(T) curves as the MST decouples above the upper critical temperature ($T_{cr}^{up}$ ~ 448 K).

In order to understand the origin of ferromagnetism in martensite phase, the magnetizing behavior at 5 K of $Mn_{1-y}Fe_yNiGe_{1-x}Si_x$ was analyzed. Here we first look back the magnetic properties of $Mn_{1-y}Fe_yNiGe$.[26] As shown in the left part of **Figure 1e**, $M_S$ increases to a plateau with a value of about 2.65 $\mu_B$ upon increasing Fe content ($y$). Fe substitution can efficiently convert AFM couplings in MnNiGe to FM ones. $H_S$ decreases monotonically and reaches a minimum of 5 kOe at $y$ = 0.24 where the AFM coupling is almost overcome by FM couplings. In $Mn_{1-y}Fe_yNiGe$ system, the high magnetization and low saturation field are achieved gradually with increasing Fe substitution. After introducing Si at Ge site, in sharp contrast, the $Mn_{1-y}Fe_yNiGe_{1-x}Si_x$ ($y$ = 0.26, 0.36, 0.46, 0.55) coheres the lowest $H_S$ and highest $M_S$ of $Mn_{1-y}Fe_yNiGe$ system and maintains them at values of $H_S$ ~ 1.35 kOe to 2.12 kOe and $M_S$ ~ 2.2 to 2.6 $\mu_B$. M(H) curves of ($y$ = 0.26, $x$) ($0.10 \leq x \leq 0.60$) series are given in **Figure 1f** (more data in Supporting Information, **Figure S4**). One can see all samples are easily magnetized in a low magnetic field and gain large saturation magnetizations of about 80 emu/g (~ 2.6 $\mu_B$). The features of low $H_S$ and high $M_S$ in $Mn_{1-y}Fe_yNiGe_{1-x}Si_x$ are highly expected to facilitate the magneto-multifunctional properties, especially the desired low-field effects.

The magnetic structure in martensite phase of $Mn_{1-y}Fe_yNiGe_{1-x}Si_x$ system was further revealed by first-principles calculations. **Figure 1g** shows the partial density of states (PDOS) of 3*d*-metal atoms in ($y$ = 0.5, $x$ = 0.5) alloy, from which one sees the remarkable spin polarizations on both Mn and Fe atoms. This accounts for the large magnetic moments of ~ 2.8 $\mu_B$ and ~ 1.8 $\mu_B$ on Mn and Fe atoms, respectively (Supporting Information, **Table S3**). Compared with Mn atom, Fe atom has a weaker polarization and a clear PDOS peak exists well below the Fermi level in spin-down state, which results in a smaller moment on Fe atom. Ni and Ge/Si atoms carry near-zero moments (**Table S3**) as the spin polarizations are very weak due to strong covalent bonds between Ni-Si/Ge atoms. **Figure 1h** depicts the spin electron density of sample ($y$ = 0.5, $x$ = 0.5), which shows the magnetization distribution in the compound. Strong localization of (positive) spin electron density values exist around both Mn/Fe atoms, indicating significant localized moments with a parallel alignment in the zigzag chains. In $Mn_{1-y}Fe_yNiGe_{1-x}Si_x$, the magnetizations, mainly originating from Mn and Fe moments with FM exchange interactions, keep high and stable values during the whole Si substitution.

In this section, we present the desired magneto-multifunctional properties in CTWs. Isothermal M(H) curves across the FMMT of sample ($y$ = 0.26, $x$ = 0.30) were measured (**Figure 2a**). As illustrated, a metamagnetic behavior is observed between 338 K with a critical field ($H_{cr}$) of about 35 kOe and 332 K with a rather low $H_{cr}$ of ~ 5 kOe. This metamagnetic behavior indicates the magnetic-field-induced martensitic structural transition from PM hexagonal parent phase to an energy-favored FM orthorhombic martensite. We further characterized the MCEs upon this FMMT. The magnetic entropy changes ($\Delta S_m$) of ($y$ = 0.26, $x$ = 0.30) alloy were derived from the loop-measured M(H) curves using Maxwell relation (See *Experimental Section*). As shown in **Figure 2b**, a value of $\Delta S_m \sim -37$ J kg$^{-1}$ K$^{-1}$ was obtained at $\Delta H \sim 50$ kOe and T ~ 335 K, showing a giant conventional (negative) MCE. Noticeably, the alloy shows a giant $\Delta S_m$ of -17 J kg$^{-1}$ K$^{-1}$ at a moderate field change of $\Delta H \sim$ 20 kOe and T ~ 333 K. These giant MCEs are benefited from the abrupt and large magnetization changes during the FMMTs as well as the low $H_S$ as discussed above.

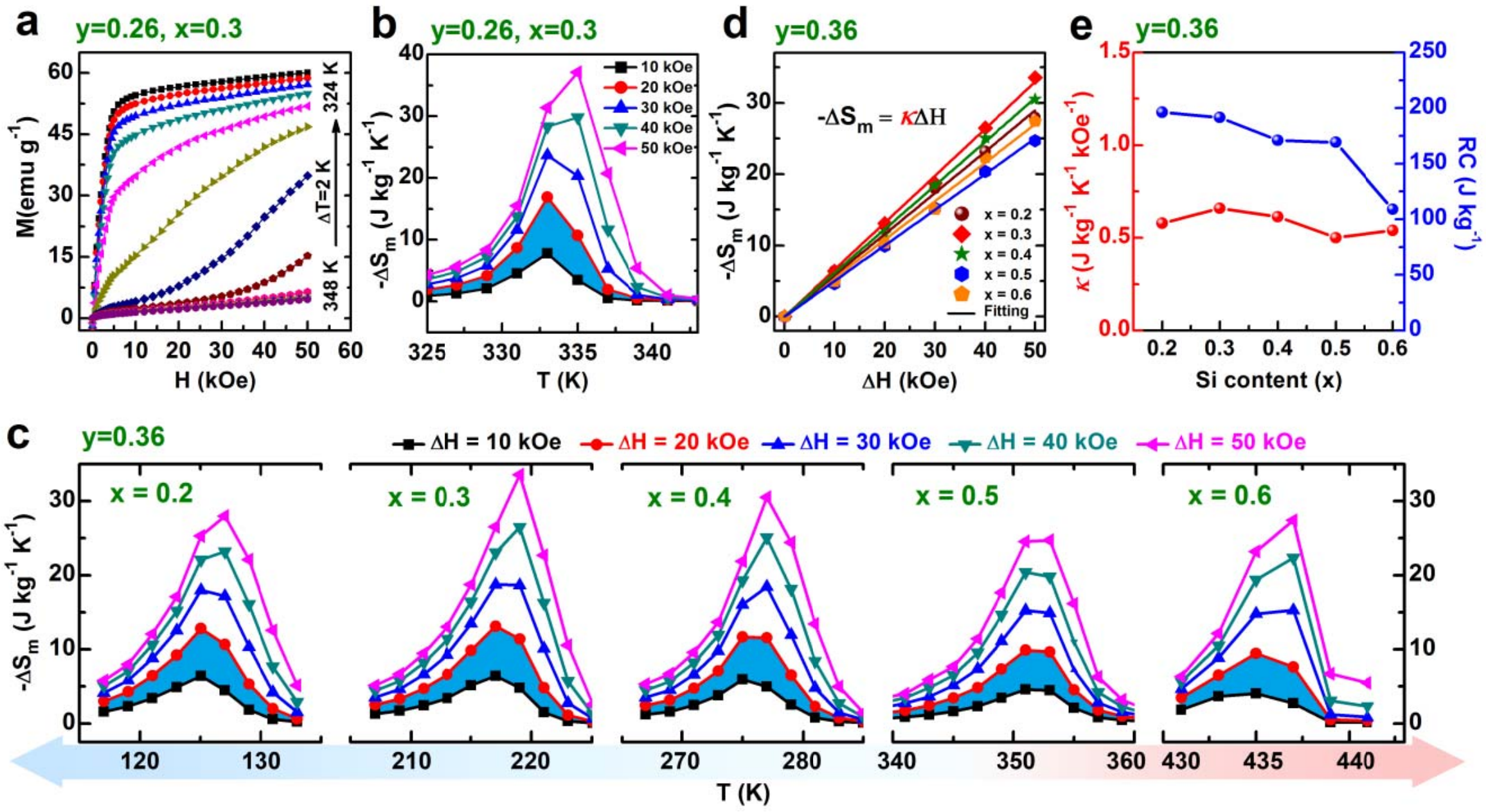

**Figure 2.** (a) Magnetic isotherms of alloy ($y$ = 0.26, $x$ = 0.3) at various temperatures around $T_t$. (b) Isothermal magnetic entropy changes ($\Delta S_m$) of alloy ($y$ = 0.26, $x$ = 0.3). (c) Isothermal magnetic entropy changes ($\Delta S_m$) of ($y$ = 0.36, $x$) series. (d) Maximum of $\Delta S_m$ for ($y$ = 0.36, $x$) series as a function of magnetic field change ($\Delta H$). Symbols refer to experiment data and the corresponding linear fitting is shown as well. Fitting formula is $-\Delta S_m = \kappa \Delta H$. (e) Si-content dependence ($x$) of $\kappa$ and refrigerant capacity (RC) of ($y$ = 0.36, $x$) series.

To systematically study the dependence of MCE on the composition, we chose alloys with ($y$ = 0.36, $x$ = 0.2, 0.3, 0.4, 0.5, 0.6) in the CTW to perform the isothermal M(H) curves and calculated the corresponding $\Delta S_m$, as shown in **Figure 2c**. It is evident that $\Delta S_m(x)$ depends on temperature very weakly in the temperature range between 120 and 440 K, and takes a value of -30 J kg$^{-1}$ K$^{-1}$ at a field change of 50 kOe, and a value of -12 J kg$^{-1}$ K$^{-1}$ at a moderate field change of 20 kOe. High-temperature *giant* MCEs, are clearly obtained in these materials. Such large MCEs are rarely found above water boiling point (see refs. [39-41]). It is important to note that these giant MCEs spanning the whole CTWs are composed of the caloric effects from both the structural and magnetic transitions and more importantly two caloric effects are locked in the same sign owing to the concurrent breaking in crystallographic and magnetic symmetries. This locked effect would greatly enhance the total caloric outcome of strongly-coupled MSTs.[6, 29]

To clearly present the dependence of maximum of $-\Delta S_m$ on the $\Delta H$, we re-plotted those values of **Figure 2c** in **Figure 2d** with the corresponding fitting curves. A linear relationship between maximum of $-\Delta S_m$ and $\Delta H$ is clearly observed up to $\Delta H = 50$ kOe. This indicates that the temperature dependence of magnetization across the FMMTs is field independent,[42] which may be attributed to the specific characteristics of the fist-order transitions in these materials. The data in **Figure 2d** can be fitted to a linear dependence, $-\Delta S_m = \kappa \Delta H$, where the coefficient $\kappa$ can be considered as the factor that describes how strong the maximum of $-\Delta S_m$ depends on $\Delta H$. The values of $\kappa$ obtained from fitting are plotted in **Figure 2e**, as a function of $x$, with the integrated refrigerant capacity (RC) that measures the effectiveness of a magnetic refrigerant. The high values of $\kappa$ and RC obtained on ($y = 0.36$, $x$) series as shown in **Figure 2e**, indicate clearly the large MCEs throughout the wide CTWs.

As well known, the functional fatigue behavior is critical for practical phase-transition-based applications.[43, 44] In order to investigate the fatigue characteristics on the FMMTs, thermal cycling tests were performed on DSC for some compositions of ($y$ = 0.26, $x$ = 0.15, 0.25) and ($y$ = 0.36, $x$ = 0.60). **Figure 3a, 3b** and **3c** show cycle-dependent calorimetric curves of alloy ($y$ = 0.26, $x$ = 0.15). Exothermic (endothermic) peak on cooling (heating) represents forward (reverse) martensitic transitions. During 93 cycles in this study, the transition calorimetric profiles were unchanged, particularly in the peak positions and peak heights, which demonstrates that the phase transitions possess good reversibility and stability. The same thermal-cycling experiments were carried out on other two alloys ($y$ = 0.26, $x$ = 0.25) and ($y$ = 0.36, $x$ = 0.60). The associated shift of $T_t$, absolute and normalized latent heats (ΔE) are presented in **Figure 3d** and **3e** (more data in Supporting Information, **Table S4**). For the transitions at low (218 K) and room (299 K) temperatures ($y$ = 0.26, $x$ = 0.15, 0.25), $T_t$ keeps almost unchanged throughout the whole cycling process with a small fluctuation less than 0.5 K (for sample with $x$ = 0.25, the shift of $T_t$ ~ 0.1 K after 93 cycles), indicating an even better functional stability than the excellent Ni-Ti-Cu/Pd alloys.[43, 44] This stability is further confirmed by a slight decay (3% ~ 5%) in ΔE following by a saturation tendency. Even for the transitions at 420 K (147 °C) in alloy ($y$ = 0.36, $x$ = 0.60), only very slight changes of $T_t$ (~ 0.8 K) and ΔE (~ 7%) can be observed. These slightly enlarged changes may be ascribed to the atom motions by high-temperature thermal activation. It is clear that the studied alloys show a potentially robust functional stability even at high temperatures, which will benefit the multifunctional applications of the materials.

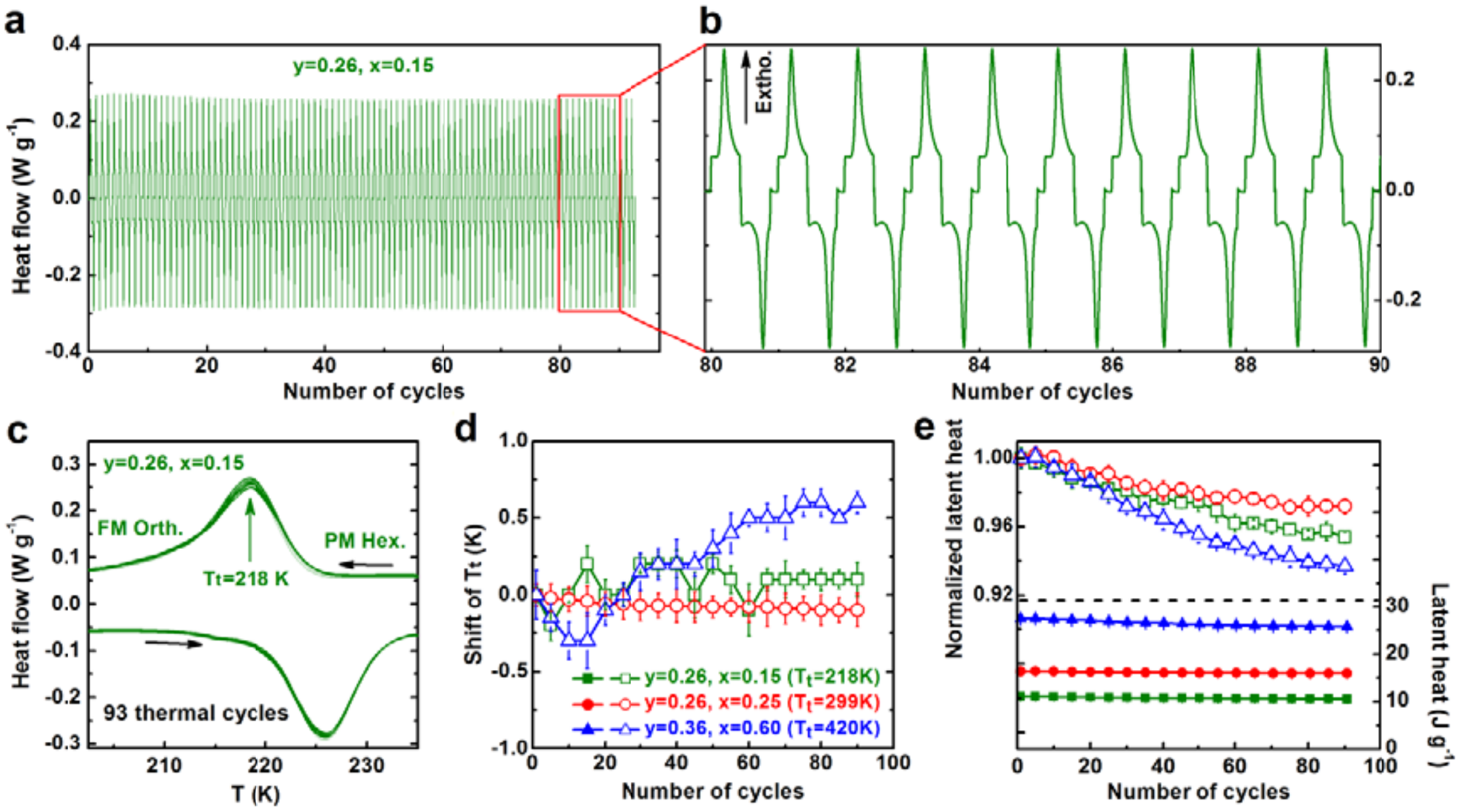

**Figure 3.** Functional fatigue behavior of $Mn_{1-y}Fe_yNiGe_{1-x}Si_x$ ($y$, $x$) system. (a), (b) and (c) DSC data of thermal cycling with 93 times around $T_t$ of alloy ($y$ = 0.26, $x$ = 0.15). An enlarged image of cycles from 80 to 89 is shown in (b). (d) Shift of $T_t$, (e) absolute (lower panel) and normalized (upper panel) latent heats (ΔE) upon cooling of alloys ($y$ = 0.26, $x$ = 0.15), ($y$ = 0.26, $x$ = 0.25) and ($y$ = 0.36, $x$ = 0.60) deprived from DSC data.

**Figure 4** shows a general comparison of the maximum entropy changes ($|\Delta S|_{max}$) of our materials and several well-known MCE materials based on current statistical data (Supporting Information, **Table S5**). The graphic is in general consistent with the one reported by Franco *et al.*[14] In this graphic, low- and room-temperature MCEs are observed in many rare-earth and Mn-based compounds, including the first-order FMMT materials such as Heusler and MM'X alloys. However, the MCEs in each family (usually composed of several material members) are limited in narrow temperature spans since many MSTs can only be tuned in a limited temperature range. It can be seen that different families show their dissimilar transition temperatures but most giant MCEs occur below 370 K. In particular, the MCEs of MM'X alloys also exist mainly in a relatively narrow range around room temperature. In sharp contrast, our $Mn_{1-y}Fe_yNiGe_{1-x}Si_x$ single material system strikingly shows giant MCEs over the very wide temperature range from 40 to 450 K. Within these wide CTWs, the magneto-strain-based applications in forms of bulks, particles or composites can also be designed. It is further conceivable that large MCEs and barocaloric effects tuned by applying pressure[18, 34] and table-like MCEs produced by composition-gradient composites[45] can be freely designed at any desired temperature within such broad range. In addition to the use as magnetic refrigerant, the giant MCE materials can also be applied as promising working substances for magnetic heat pumps or electric power generations.[12, 20] For these applications, the MSTs are required to occur at high temperatures in order to gain the heats efficiently and cyclically from ambient environments, driven by hot water or vapor from sea surface (30 ~ 40 °C) or geothermal spring (50 ~ 100 °C), or by the exhausted heats from applications of hypercritical $CO_2$ (~ 120 °C), automobile and industrial production (100 ~ 500 °C). For the high-temperature MSTs between 373 and 448 K, furthermore, the corresponding materials show an increasingly low cost due to the high-level Si-substitution for expensive Ge element.

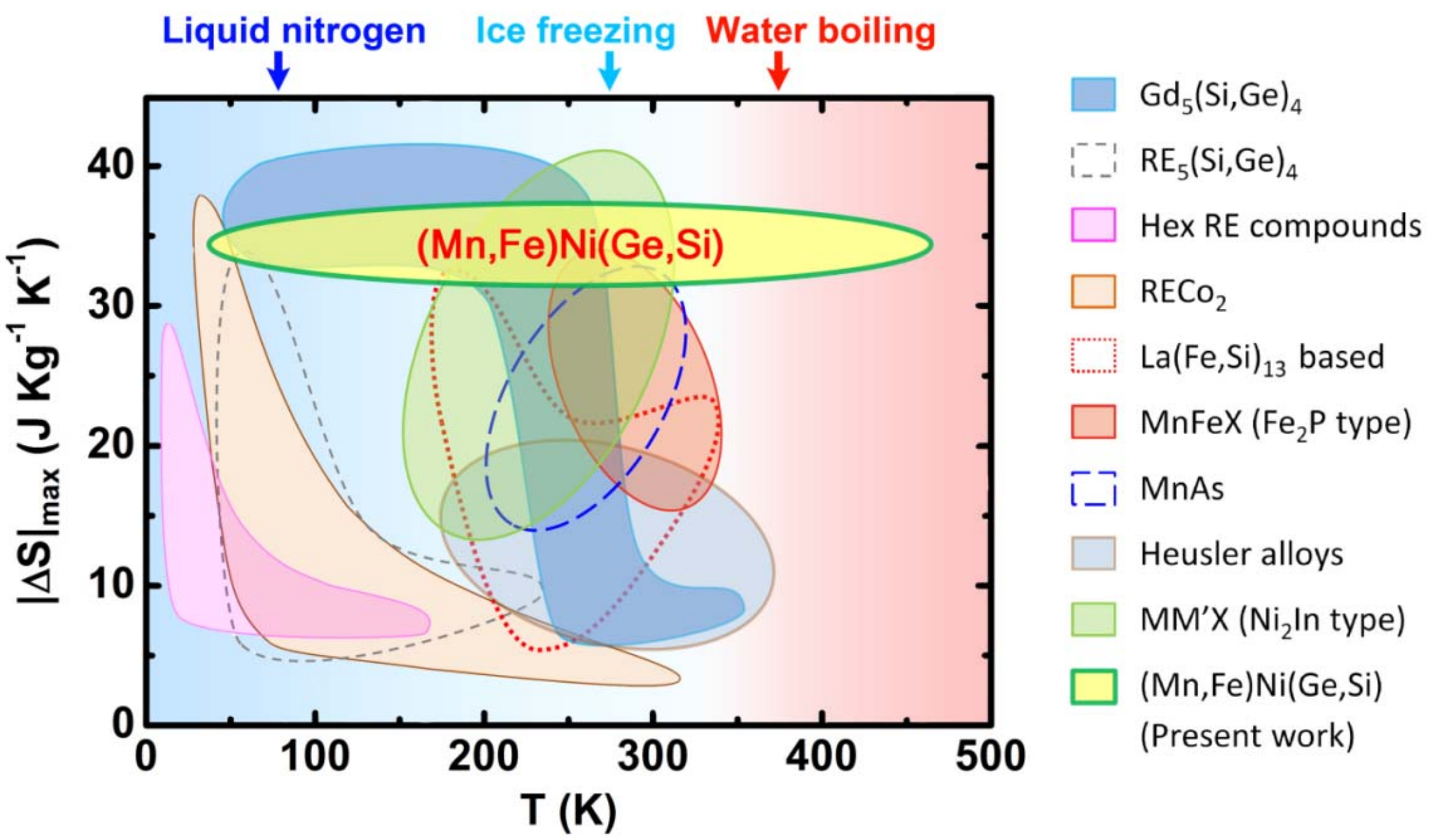

**Figure 4**. A statistical graphic of absolute values of maximum entropy changes ($|\Delta S|_{max}$) for ΔH = 50 kOe versus peak temperature for different families of magnetocaloric materials, including the present (Mn,Fe)Ni(Ge,Si) system. Detail data and relevant references are included in **Supporting Information**, **Table S5**. Here, RE refers to rare-earth elements, Hex to hexagonal structure, and X to *p*-block elements.

In conclusion, applying the principle of isostructural alloying, we have successfully realized the strongly-coupled magnetostructural transitions within a series of unprecedentedly wide Curie-temperature windows, with widths as large as 400 K, in a single host system (Mn,Fe)Ni(Ge,Si). This work reveals clear physical pictures of the tuning route in material design and the Curie-temperature windows for phase transitions. The combination of low-field large MCEs, locked caloric effects, wide working temperature, and robust functional stability makes the materials promising for various smart applications including shape-memory/strain-based output,[46] solid-state multi-caloric cooling/heating[19] and energy conversion.[20] The unprecedentedly wide Curie-temperature windows provide a broad design platform of magnetostructural transitions for tunable magneto-multifunctional properties of the rare-earth-free Mn-based (Mn,Fe)Ni(Ge,Si) multiferroic magnetoelastic alloys. The strongly-coupled magnetostructural transitions can be manipulated in frameworks of compositional engineering,[29, 47] pressure modulations,[18, 34, 48] and low-dimensional forming[4, 49] for potential applications, including strain-based composites,[50, 51] functional devices,[46, 52] and multiferroic heterostructures.[53, 54]

## Experimental Section

*Experimental Methods:* Highly purified metals were arc-melted under argon atmosphere. Each ingot was melted for four times and was turned over after each melting process. As-cast ingots were annealed in evacuated quartz tubes filled with high-purity argon at 1123 K for 5 days and cooled down to room temperature slowly. Powder x-ray diffraction (XRD) analysis were performed on a Rigaku D/max 2,400 X-ray diffractometer with Cu-$K\alpha$ radiation at room temperature. Magnetic measurements including thermomagnetic curves and isothermal magnetization curve, were carried out on a superconductive interference device (SQUID) magnetometer and physical property measurement system (PPMS, Quantum Design). Thermomagnetic curves with temperature above 400 K were performed on a PPMS plus vibrating sample magnetometer (VSM) with a maximum field of 50 kOe. Thermal analysis were performed using differential scanning calorimetry (DSC STA 449 F3 Jupiter and DSC 214 Ployma, NETZSCH). The ramping rate is 5 K $min^{-1}$. By adding an apparatus involving two permanent magnets, the gravity change caused by magnetic transition can be detected on the thermogravimetric analyzer. The magnetic entropy changes ($\Delta S_m$) across magnetostructural transitions were derived from the magnetization curves using the Maxwell relation[14]: $\Delta S_m(T,H) = S_m(T,H) - S_m(T,0) = \int_0^H \left(\frac{\partial M}{\partial T}\right)_H dH$. The RC is defined as $RC = \int_{T_1}^{T_2} |\Delta S_m| \, dT$, where $T_1$ and $T_2$ are two temperatures of the half maximum of $\Delta S_m(T)$ peak.[14] To avoid fake $\Delta S_m$ values, the temperature loop process method[55] was applied to measure the isothermal magnetization curves with an interval of 2 K for these FMMTs with a clear thermal hysteresis.

*Calculation Methods*: The first-principles calculations on density of states (DOS), magnetic moment distribution and spin electron density (SED) were performed using ultrasoft pseudopotential method based on the density function theory (DFT).[56] The exchange and correlation energy were treated using the generalized gradient approximations. 500-eV cutoff-energy in the plane-wave basis and 120 *k*-points in the irreducible Brillouin zone were used for a good convergence of the total energy. The total-energy difference tolerance for the self-consistent field iteration was set at $1\times10^{-6}$ eV atom$^{-1}$.

## Acknowledgements

This work was supported by National Natural Science Foundation of China (51301195, 11174352, and 51431009), National Basic Research Program of China (2012CB619405 and 2011CB012800), Beijing Municipal Science & Technology Commission (Z141100004214004), and Youth Innovation Promotion Association of CAS (2013002).

## References

[1] W. Eerenstein, N. D. Mathur, J. F. Scott, *Nature* **2006**, *442*, 759.

[2] E. K. H. Salje, *Annu. Rev. Mater. Res.* **2012**, *42*, 265.

[3] R. Kainuma, Y. Imano, W. Ito, Y. Sutou, H. Morito, S. Okamoto, O. Kitakami, K. Oikawa, A. Fujita, T. Kanomata, K. Ishida, *Nature* **2006**, *439*, 957.

[4] M. Chmielus, X. X. Zhang, C. Witherspoon, D. C. Dunand, P. Mullner, *Nat. Mater.* **2009**, *8*, 863.

[5] T. Krenke, E. Duman, M. Acet, E. F. Wassermann, X. Moya, L. Manosa, A. Planes, *Nat. Mater.* **2005**, *4*, 450.

[6] J. Liu, T. Gottschall, K. P. Skokov, J. D. Moore, O. Gutfleisch, *Nat. Mater.* **2012**, *11*, 620.

[7] S. Singh, R. Rawat, S. E. Muthu, S. W. D'Souza, E. Suard, A. Senyshyn, S. Banik, P. Rajput, S. Bhardwaj, A. M. Awasthi, R. Ranjan, S. Arumugam, D. L. Schlagel, T. A. Lograsso, A. Chakrabarti, S. R. Barman, *Phys. Rev. Lett.* **2012**, *109*, 246601.

[8] A. K. Nayak, M. Nicklas, S. Chadov, C. Shekhar, Y. Skourski, J. Winterlik, C. Felser, *Phys. Rev. Lett.* **2013**, *110*, 127204.

[9] H. E. Karaca, I. Karaman, B. Basaran, Y. Ren, Y. I. Chumlyakov, H. J. Maier, *Adv. Funct. Mater.* **2009**, *19*, 983.

[10] N. Sarawate, M. Dapino, *Appl. Phys. Lett.* **2006**, *88*, 121923.

[11] S. Fähler, U. K. Rößler, O. Kastner, J. Eckert, G. Eggeler, H. Emmerich, P. Entel, S.

Müller, E. Quandt, K. Albe, *Adv. Eng. Mater.* **2012**, *14*, 10.

[12] B. Yu, M. Liu, P. W. Egolf, A. Kitanovski, *Int. J. Refrig.* **2010**, *33*, 1029.

[13] O. Gutfleisch, M. A. Willard, E. Brück, C. H. Chen, S. G. Sankar, J. P. Liu, *Adv. Mater.* **2011**, *23*, 821.

[14] V. Franco, J. S. Blazquez, B. Ingale, A. Conde, *Annu. Rev. Mater. Res.* **2012**, *42*, 305.

[15] B. Li, W. J. Ren, Q. Zhang, X. K. Lv, X. G. Liu, H. Meng, J. Li, D. Li, Z. D. Zhang, *Appl. Phys. Lett.* **2009**, *95*, 172506.

[16] T. Kihara, X. Xu, W. Ito, R. Kainuma, M. Tokunaga, *Phys. Rev. B* **2014**, *90*, 214409.

[17] A. Planes, L. Manosa, M. Acet, *J. Phys. - Cond. Matter* **2009**, *21*, 233201.

[18] L. Manosa, D. González-Alonso, A. Planes, E. Bonnot, M. Barrio, J. L. Tamarit, S. Aksoy, M. Acet, *Nat. Mater.* **2010**, *9*, 478.

[19] X. Moya, S. Kar-Narayan, N. D. Mathur, *Nat. Mater.* **2014**, *13*, 439.

[20] V. Srivastava, Y. Song, K. Bhatti, R. D. James, *Adv. Energy Mater.* **2011**, *1*, 97.

[21] Y. T. Song, K. P. Bhatti, V. Srivastava, C. Leighton, R. D. James, *Energy Environ. Sci.* **2013**, *6*, 1315.

[22] S. Y. Yu, L. Ma, G. D. Liu, Z. H. Liu, J. L. Chen, Z. X. Cao, G. H. Wu, B. Zhang, X. X. Zhang, *Appl. Phys. Lett.* **2007**, *90*, 242501.

[23] V. K. Pecharsky, K. A. Gschneidner, *Appl. Phys. Lett.* **1997**, *70*, 3299.

[24] O. Tegus, E. Bruck, L. Zhang, Dagula, K. H. J. Buschow, F. R. de Boer, *Physica B* **2002**, *319*, 174.

[25] A. Barcza, Z. Gercsi, K. S. Knight, K. G. Sandeman, *Phys. Rev. Lett.* **2010**, *104*, 247202.

[26] E. K. Liu, W. H. Wang, L. Feng, W. Zhu, G. J. Li, J. L. Chen, H. W. Zhang, G. H. Wu, C. B. Jiang, H. B. Xu, F. de Boer, *Nat. Commun.* **2012**, *3*, 873.

[27] D. Choudhury, T. Suzuki, Y. Tokura, Y. Taguchi, *Sci. Rep.* **2014**, *4*, 7544.

[28] T. Samanta, I. Dubenko, A. Quetz, S. Stadler, N. Ali, *Appl. Phys. Lett.* **2012**, *101*, 242405.

[29] E. K. Liu, H. G. Zhang, G. Z. Xu, X. M. Zhang, R. S. Ma, W. H. Wang, J. L. Chen, H. W. Zhang, G. H. Wu, L. Feng, X. X. Zhang, *Appl. Phys. Lett.* **2013**, *102*, 122405.

[30] C. L. Zhang, D. H. Wang, Z. D. Han, B. Qian, H. F. Shi, C. Zhu, J. Chen, T. Z. Wang, *Appl. Phys. Lett.* **2013**, *103*, 132411.

[31] G. J. Li, E. K. Liu, H. G. Zhang, Y. J. Zhang, J. L. Chen, W. H. Wang, H. W. Zhang, G. H. Wu, S. Y. Yu, *J. Magn. Magn. Mater.* **2013**, *332*, 146.

[32] J. H. Chen, E. K. Liu, X. Qi, H. Z. Luo, W. H. Wang, H. W. Zhang, S. G. Wang, J. W. Cai, G. H. Wu, *Comp. Mater. Sci.* **2014**, *89*, 130.

[33] L. F. Zhang, J. M. Wang, H. Hua, C. B. Jiang, H. B. Xu, *Appl. Phys. Lett.* **2014**, *105*, 112402.

[34] T. Samanta, D. L. Lepkowski, A. U. Saleheen, A. Shankar, J. Prestigiacomo, I. Dubenko, A. Quetz, I. W. H. Oswald, G. T. McCandless, J. Y. Chan, P. W. Adams, D. P. Young, N. Ali, S. Stadler, *Phys. Rev. B* **2015**, *91*, 020401.

[35] T. Samanta, D. L. Lepkowski, A. U. Saleheen, A. Shankar, J. Prestigiacomo, I. Dubenko, A. Quetz, I. W. H. Oswald, G. T. McCandless, J. Y. Chan, P. W. Adams, D. P. Young, N. Ali, S. Stadler, *J. Appl. Phys.* **2015**, *117*, 123911.

[36] C. L. Zhang, H. F. Shi, E. J. Ye, Y. G. Nie, Z. D. Han, D. H. Wang, *J. Alloys Compd.* **2015**, *639*, 36.

[37] V. Johnson, *Inorg. Chem.* **1975**, *14*, 1117.

[38] W. Bażela, A. Szytuła, J. Todorovic, A. Zieba, *Phys. Status Solidi A* **1981**, *64*, 367.

[39] V. Franco, J. S. Blázquez, C. F. Conde, A. Conde, *Appl. Phys. Lett.* **2006**, *88*, 042505.

[40] M. Jasinski, J. Liu, S. Jacobs, C. Zimm, *J. Appl. Phys.* **2010**, *107*, 09A953.

[41] W. J. Feng, Q. Zhang, L. Q. Zhang, B. Li, J. Du, Y. F. Deng, Z. D. Zhang, *Solid State Commun.* **2010**, *150*, 949.

[42] V. Franco, A. Conde, *Int. J. Refrig.* **2010**, *33*, 465.

[43] R. Zarnetta, R. Takahashi, M. L. Young, A. Savan, Y. Furuya, S. Thienhaus, B. Maaß, M. Rahim, J. Frenzel, H. Brunken, Y. S. Chu, V. Srivastava, R. D. James, I. Takeuchi, G. Eggeler, A. Ludwig, *Adv. Funct. Mater.* **2010**, *20*, 1917.

[44] Y. T. Song, X. Chen, V. Dabade, T. W. Shield, R. D. James, *Nature* **2013**, *502*, 85.

[45] L. Li, M. Kadonaga, D. Huo, Z. Qian, T. Namiki, K. Nishimura, *Appl. Phys. Lett.* **2012**, *101*, 122401.

[46] J. Liu, N. Scheerbaum, S. Kauffmann-Weiss, O. Gutfleisch, *Adv. Eng. Mater.* **2012**, *14*, 653.

[47] I. Takeuchi, O. O. Famodu, J. C. Read, M. A. Aronova, K. S. Chang, C. Craciunescu, S.

E. Lofland, M. Wuttig, F. C. Wellstood, L. Knauss, A. Orozco, *Nat. Mater.* **2003**, *2*, 180.

[48] L. Caron, N. T. Trung, E. Brück, *Phys. Rev. B* **2011**, *84*, 020414.

[49] J. S. Juan, M. L. No, C. A. Schuh, *Nat. Nano.* **2009**, *4*, 415.

[50] D. C. Hofmann, *Science* **2010**, *329*, 1294.

[51] Y. Y. Zhao, F. X. Hu, L. F. Bao, J. Wang, H. Wu, Q. Z. Huang, R. R. Wu, Y. Liu, F. R. Shen, H. Kuang, M. Zhang, W. L. Zuo, X. Q. Zheng, J. R. Sun, B. G. Shen, *J. Am. Chem. Soc.* **2015**, *137*, 1746.

[52] M. Gueltig, H. Ossmer, M. Ohtsuka, H. Miki, K. Tsuchiya, T. Takagi, M. Kohl, *Adv. Energy Mater.* **2014**, *4*, 1400751.

[53] X. Moya, L. E. Hueso, F. Maccherozzi, A. I. Tovstolytkin, D. I. Podyalovskii, C. Ducati, L. C. Phillips, M. Ghidini, O. Hovorka, A. Berger, M. E. Vickers, E. Defay, S. S. Dhesi, N. D. Mathur, *Nat. Mater.* **2013**, *12*, 52.

[54] Y. Y. Gong, D. H. Wang, Q. Q. Cao, E. K. Liu, J. Liu, Y. W. Du, *Adv. Mater.* **2015**, *27*, 801.

[55] L. Caron, Z. Q. Ou, T. T. Nguyen, D. T. C. Thanh, O. Tegus, E. Brück, *J. Magn. Magn. Mater.* **2009**, *321*, 3559.

[56] M. C. Payne, M. P. Teter, D. C. Allan, T. A. Arias, J. D. Joannopoulos, *Rev. Mod. Phys.* **1992**, *64*, 1045.

Supporting Information

# Unprecedentedly Wide Curie-Temperature Windows as Phase-Transition Design Platform for Tunable Magneto-Multifunctional Materials

*Zhi-Yang Wei, En-Ke Liu,* Yong Li, Gui-Zhou Xu, Xiao-Ming Zhang, Guo-Dong Liu, Xue-Kui Xi, Hong-Wei Zhang, Wen-Hong Wang,* Guang-Heng Wu, and Xi-Xiang Zhang*

**X-ray diffraction (XRD) structural analysis.** The powder X-ray diffraction measurements for $y$ = 0.26, 0.36, 0.46 and 0.55 with various $x$ in $Mn_{1-y}Fe_yNiGe_{1-x}Si_x$ system were performed on well-grinded powders. Upon the measurements, the scanning speed was set as 5 degree per minute. And the diffraction peaks from Cu K$\alpha_2$ radiation were carefully deducted. The data are shown in **Figure S1** below**.** Good diffraction quality was obtained for the samples with low-Si substitution to high-Si substitution for $y$ = 0.26, 0.36, 0.46 and 0.55. The $Ni_2In$-type hexagonal parent structure and TiNiSi-type martensite structure are well identified in the diffraction patterns. The Miller indices $(hkl)_h$ and $(hkl)_o$ for parent and martensite structures were indexed. The results indicate that $Mn_{1-y}Fe_yNiGe_{1-x}Si_x$ alloys can well crystallize the desired phases for the strong degree of substitution with $y$ = 0.26, 0.36, 0.46, 0.55 and $0 \leq x \leq 1$, which provides a critical fundament for the realization of our strongly-coupled magnetostructural transitions in an extremely wide temperature range.

After careful examinations, nevertheless, two extra small diffraction peaks were observed at around 37.6° and 45.4° in the XRD patterns of some alloy compositions (as pointed out with ♦ in **Figure S1**). In the present work, the samples were prepared by slow cooling with furnace after annealing, in order to get the equilibrium phase transitions without stress and disorders in samples. This impurity phase with very small amounts probably result from the precipitate during cooling process. Nevertheless, from the diffraction intensity the phase proportion of this impurity phase is rather low. One can see that, in general, this impurity phase imposes little influence on phase transitions and magnetic properties of the samples, which has also been evidenced by the regularity of Si-content dependence of both martensitic transition temperature and Curie temperature shown in the phase diagram and the M-T curves in **Figure 1**. Importantly, the highly tunable phase transitions and giant magnetocaloric effects were obtained in such an unprecedentedly wide temperature range. It is hopeful to avoid this small-amount impurity phase by quenching the samples after annealing. Further optimizations and enhancements are expected in the important physical properties obtained in present work.

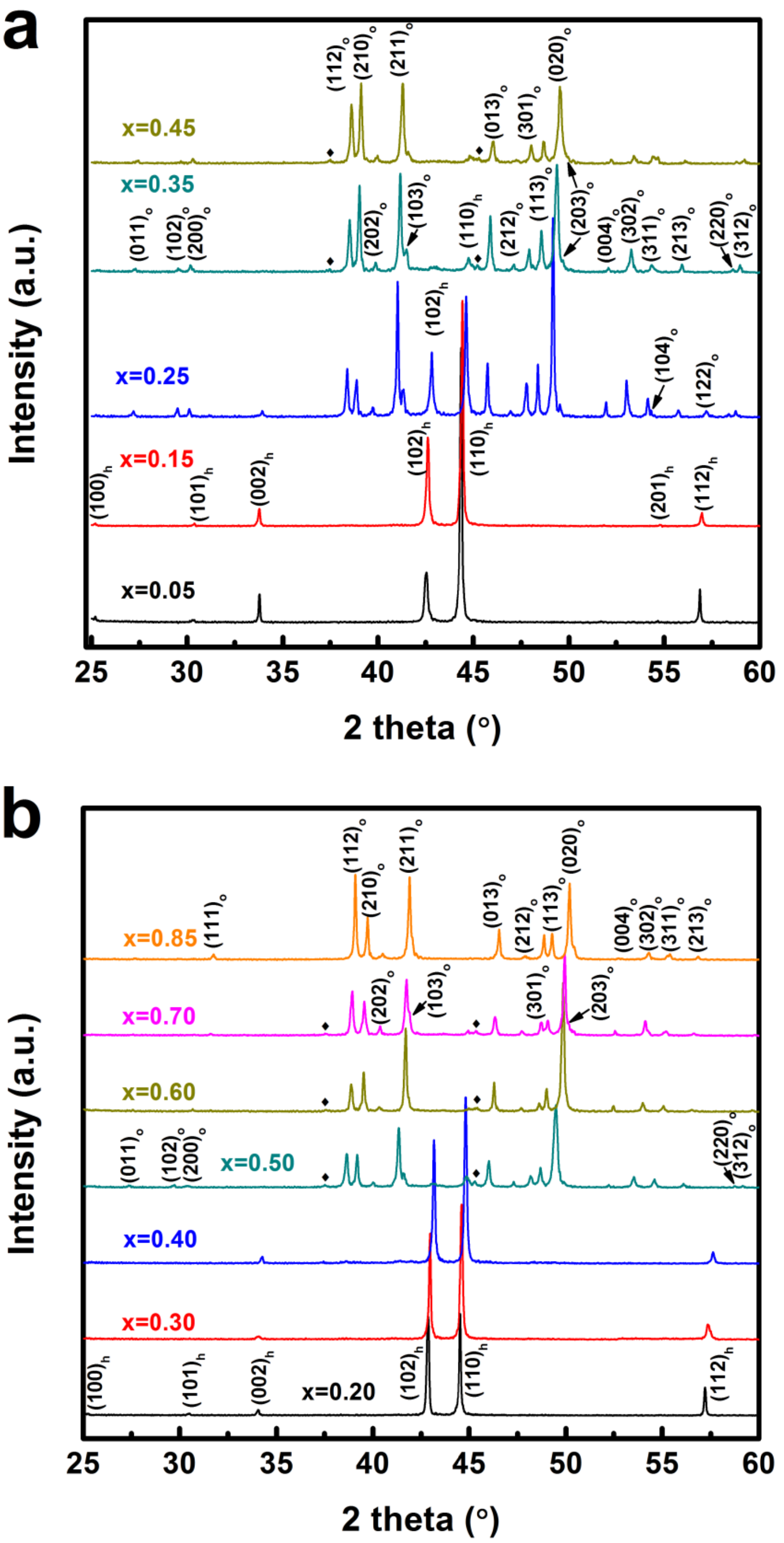

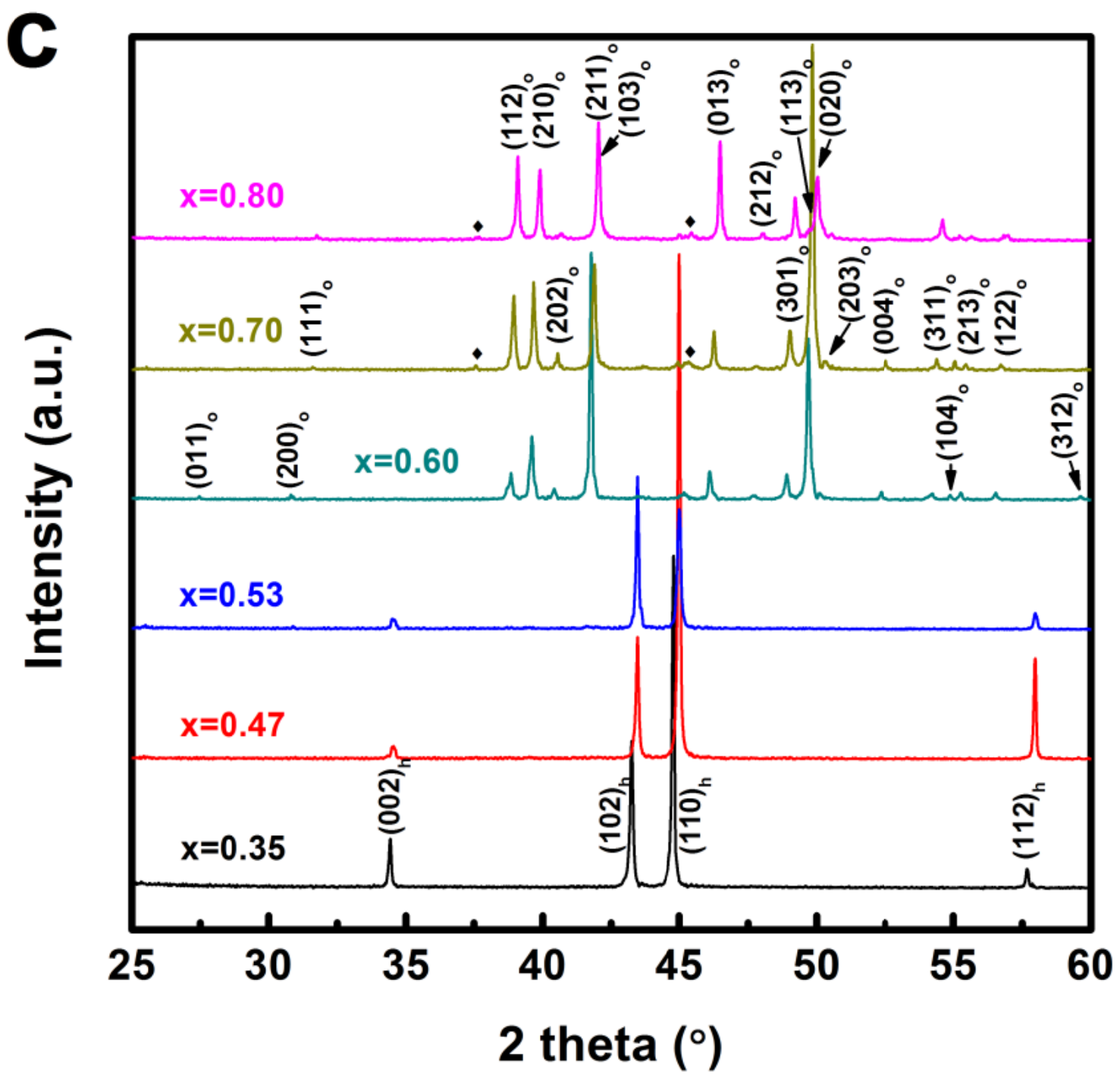

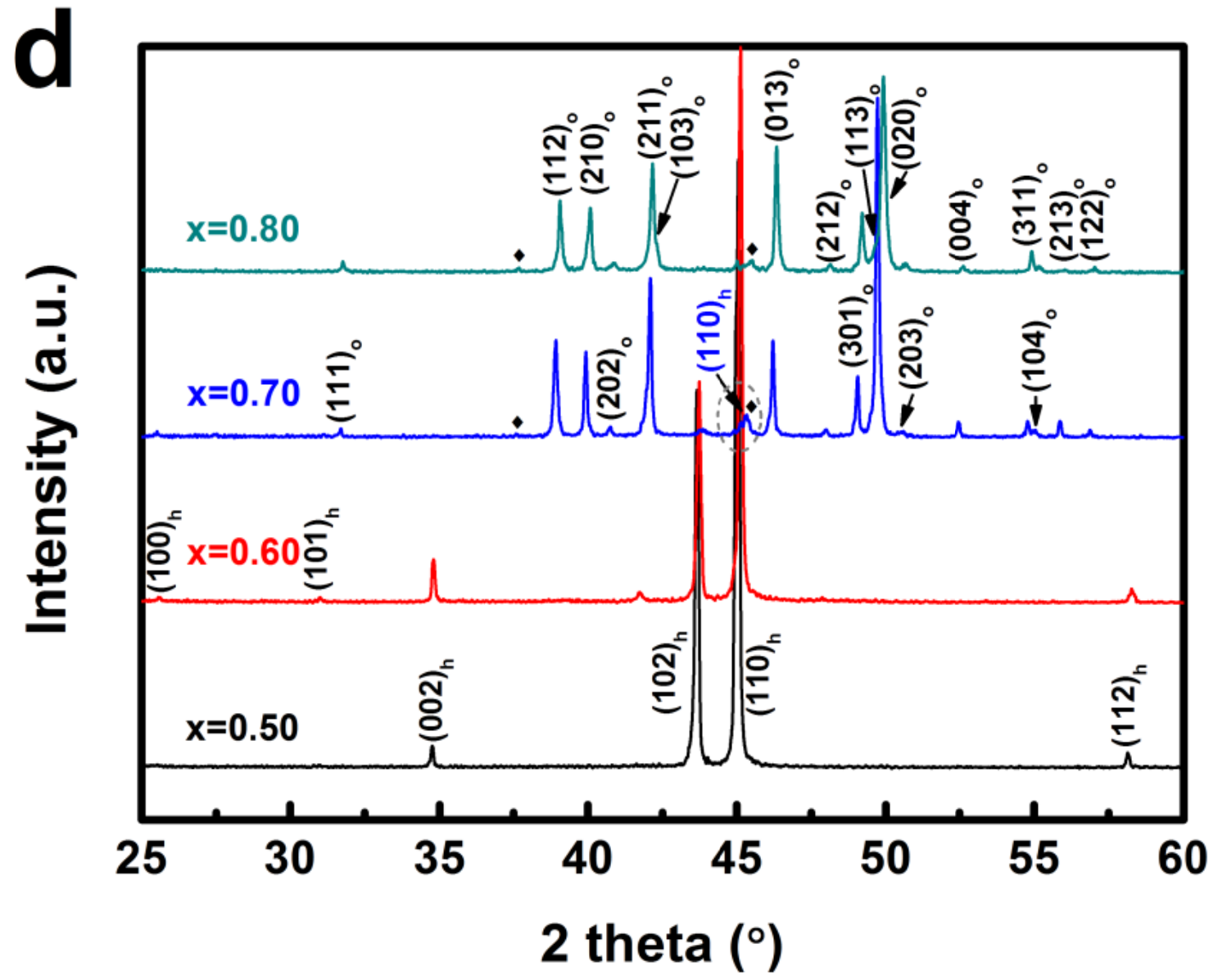

**Figure S1.** XRD patterns with various $x$ for (a) $Mn_{0.74}Fe_{0.26}NiGe_{1-x}Si_x$, (b) $Mn_{0.64}Fe_{0.36}NiGe_{1-x}Si_x$, (c) $Mn_{0.54}Fe_{0.46}NiGe_{1-x}Si_x$, (d) $Mn_{0.45}Fe_{0.55}NiGe_{1-x}Si_x$.

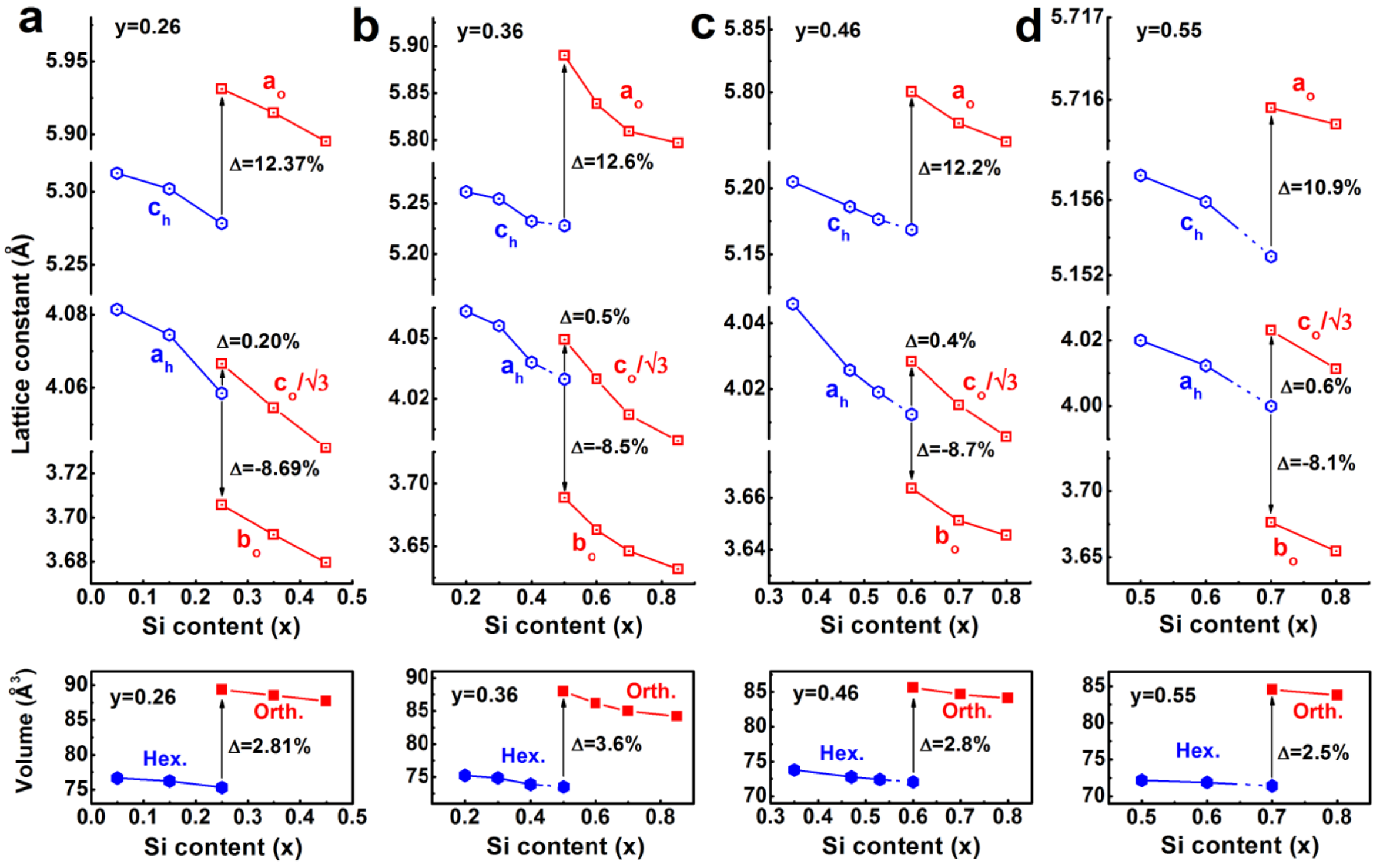

**Figure S2. Si-content dependence of lattice constant and cell volume across the with martensitic transitions for (a) $Mn_{0.74}Fe_{0.26}NiGe_{1-x}Si_x$, (b) $Mn_{0.64}Fe_{0.36}NiGe_{1-x}Si_x$, (c) $Mn_{0.54}Fe_{0.46}NiGe_{1-x}Si_x$, (d) $Mn_{0.45}Fe_{0.55}NiGe_{1-x}Si_x$. The hexagonal parent and orthorhombic martensite phase are represented by subscript "h" and "o", respectively.**

**Lattice constant and cell volume across the martensitic transitions.** The lattice constants of $Mn_{1-y}Fe_yNiGe_{1-x}Si_x$ alloys are deprived from fitting the XRD patterns in **Figure S1**, as depicted in **Figure S2**. The dash lines refer to the estimated data points. From the results, the lattice constants for both the parent and martensite phases decrease with increasing Si content for various $y$ values, which can be ascribed to the smaller atom size of Si compared to Ge. The axes and volumes of the two structures are related as $a_o = c_h$, $b_o = a_h$, $c_o=\sqrt{3}a_h$, and $V_o = 2V_h$, and the distortion way of the lattice when the martensitic transition occurs are indicated by the arrows. Upon the martensitic transition, the extension along $c_h$ reaches 12.37%, and the volume expansion is as large as 2.81% for $y$ = 0.26. The detailed data of the lattice constants and volumes are shown in **Table S1**.

**Table S1. Lattice constants and cell volumes of parent and martensite phases for $Mn_{1-y}Fe_yNiGe_{1-x}Si_x$ alloys.**

| | $x$ | $a_h$ [Å] | $c_h$ [Å] | $c_h/a_h$ | $V_h$ [Å$^3$] | $a_o$ [Å] | $b_o$ [Å] | $c_o$ [Å] | $V_o$ [Å$^3$] | $\Delta V_{o-h}$** [%] |
|---|---|---|---|---|---|---|---|---|---|---|
| | 0.05 | 4.0814 | 5.3125 | 1.3016 | 76.6373 | - | - | - | - | - |
| | 0.15 | 4.0745 | 5.3020 | 1.3013 | 76.2288 | - | - | - | - | - |
| $y$=0.26 | 0.25 | 4.0585 | 5.2782 | 1.3005 | 75.2918 | 5.9311 | 3.7058 | 7.0434 | 77.4045 | 2.81 |
| | 0.35 | - | - | - | - | 5.9148 | 3.6923 | 7.0226 | 77.4045 | - |
| | 0.45 | - | - | - | - | 5.8952 | 3.6796 | 7.0034 | 76.6834 | - |
| | 0.2 | 4.0633 | 5.2614 | 1.2949 | 75.2289 | - | - | - | - | - |
| | 0.3 | 4.0561 | 5.2544 | 1.2954 | 74.8637 | - | - | - | - | - |
| | 0.4 | 4.0380 | 5.2323 | 1.2958 | 73.8849 | - | - | - | - | - |
| $y$=0.36 | 0.5 | 4.030* | 5.228* | 1.297* | 73.521* | 5.8898 | 3.6886 | 7.0134 | 76.1850 | 3.6 |
| | 0.6 | - | - | - | - | 5.8386 | 3.6629 | 6.9798 | 74.6362 | - |
| | 0.7 | - | - | - | - | 5.8093 | 3.6462 | 6.9491 | 73.5958 | - |
| | 0.85 | - | - | - | - | 5.7971 | 3.6319 | 6.9269 | 72.9214 | - |
| | 0.35 | 4.0459 | 5.2051 | 1.2865 | 73.7887 | - | - | - | - | - |
| | 0.47 | 4.0258 | 5.1860 | 1.2882 | 72.7893 | - | - | - | - | - |
| $y$=0.46 | 0.53 | 4.0191 | 5.1766 | 1.2880 | 72.4158 | - | - | - | - | - |
| | 0.6 | 4.012* | 5.169* | 1.288* | 72.061* | 5.8004 | 3.6636 | 6.9774 | 74.1361 | 2.8 |
| | 0.7 | - | - | - | - | 5.7755 | 3.6513 | 6.9546 | 73.3278 | - |
| | 0.8 | - | - | - | - | 5.7605 | 3.6455 | 6.9379 | 72.8469 | - |
| | 0.5 | 4.02 | 5.1573 | 1.2829 | 72.1781 | - | - | - | - | - |
| $y$=0.55 | 0.6 | 4.0123 | 5.1559 | 1.2850 | 71.8823 | - | - | - | - | - |
| | 0.7 | 4.000* | 5.153* | 1.295* | 71.402* | 5.7159 | 3.6765 | 6.9680 | 73.2145 | 2.5 |
| | 0.8 | - | - | | - | 5.7157 | 3.6546 | 6.9477 | 72.5638 | - |

*the estimated data.

**$\Delta V_{o-h} = V_o/V_h - 1$.

**Magnetic measurements.** The thermomagnetic curves (**Figure S3**) and isothermal magnetization curves (**Figure S4**) were measured for $y$ = 0.26, 0.36, 0.46 and 0.55 with various $x$ in $Mn_{1-y}Fe_yNiGe_{1-x}Si_x$ system. The thermomagnetic curves show that the magnetostructural transition temperature can be continuously tuned within a wide temperature range.

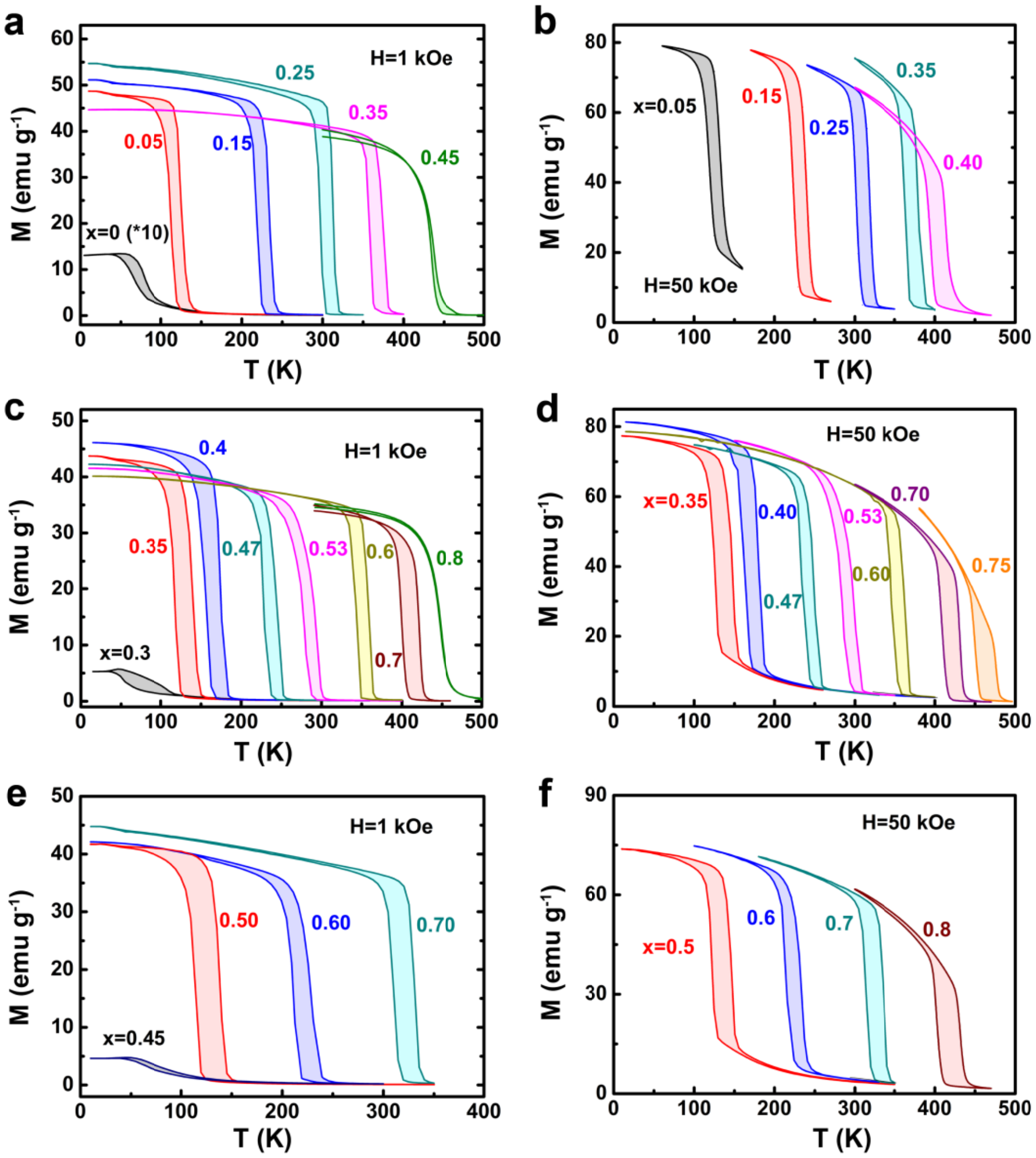

**Figure S3. Thermomagnetic curves of $Mn_{1-y}Fe_yNiGe_{1-x}Si_x$ for (a) $y$ = 0.26, (c) $y$ = 0.46, and (e) $y$ = 0.55, under magnetic field of 1 kOe, and for (b) $y$ = 0.26, (d) $y$ = 0.46, and (f) $y$ = 0.55, under magnetic field of 50 kOe.**

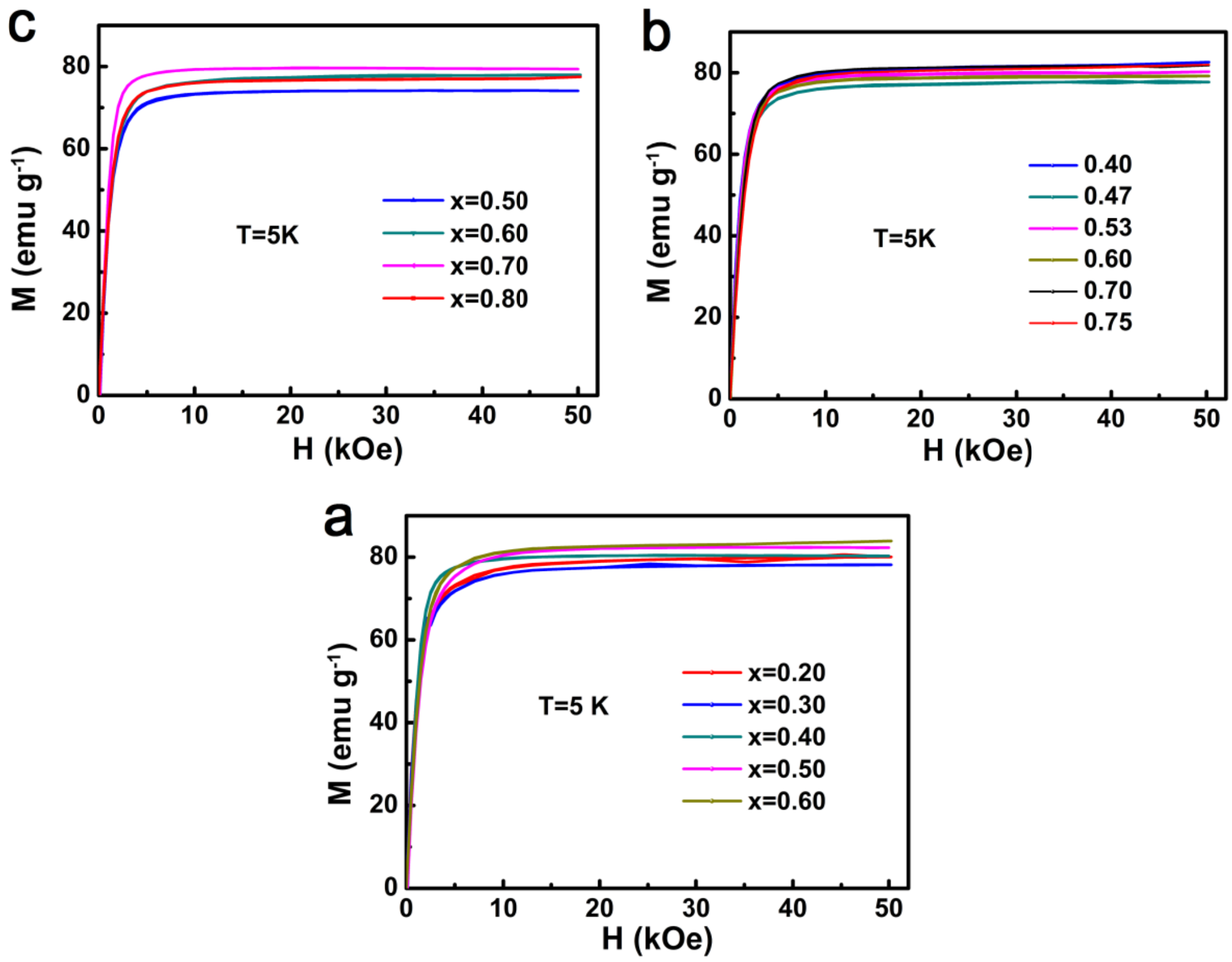

**Figure S4. Isothermal magnetization curves of $Mn_{1-y}Fe_yNiGe_{1-x}Si_x$ for (a) *y* = 0.36, (b) *y* = 0.46, and (c) *y* = 0.55, at 5 K.**

**Thermal Measurements.** Thermal analysis were performed on differential scanning calorimeter (DSC) for representative alloys with magnetostructural transition temperature from high (above room temperature, **Figure S5**) to low (**Figure S6**) temperatures. By adding an apparatus involving two permanent magnets to the high-temperature DSC, magnetic transition can be detected according to the thermal gravity (TG) change on the thermogravimetric analyzer. The big exothermic (endothermic) peak upon cooling (heating) on DSC curves indicates the (inverse) martensitic transitions. This first order phase transition shows apparent thermal hysteresis. In contrast, the rather small but observable peaks with no thermal hysteresis on DSC curves result from the magnetic transition at Curie temperature. The magnetic transitions (including magnetic ordering at Curie temperature and magnetostructural transitions) can be also revealed by the abrupt TG signal changes on corresponding TG curve. For a magnetostructural transition, the coupling of martensitic and magnetic transitions leads to a simultaneity between big exothermic peak on DSC curve and gravity change on TG curve upon cooling, as instance for the alloy (*y* = 0.36, *x* = 0.60) in **Figure S5**.

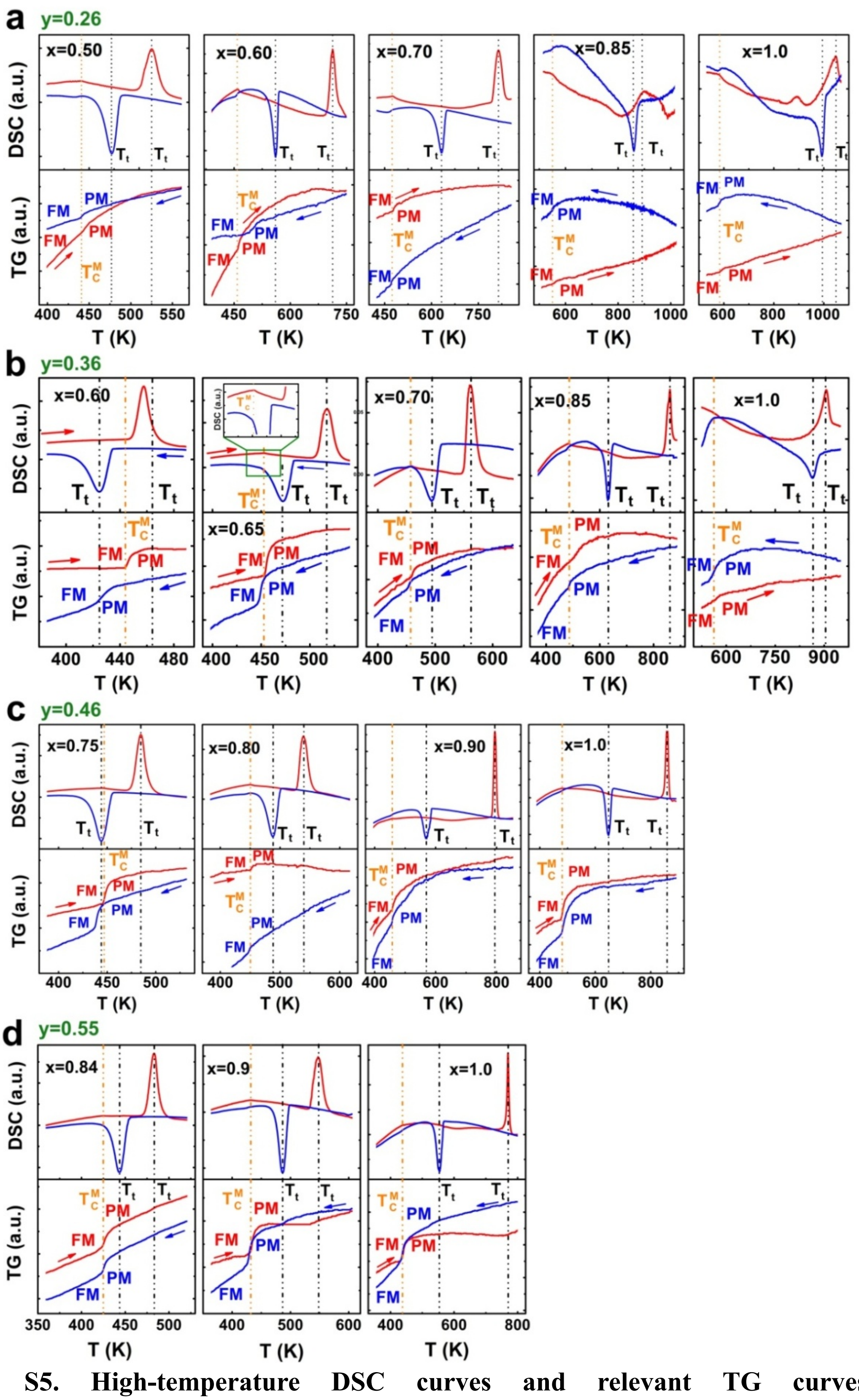

**Figure S5. High-temperature DSC curves and relevant TG curves for $Mn_{1-y}Fe_yNiGe_{1-x}Si_x$ for (a) $y$ = 0.26, (b) $y$ = 0.36, (c) $y$ = 0.46 and (d) $y$ = 0.55. FM, PM, $T_t$ and $T_C^M$ refer to ferromagnetic, paramagnetic, martensitic transition temperature and Curie temperature of martensite, respectively. The red and blue lines correspond to heating and cooling process respectively.**

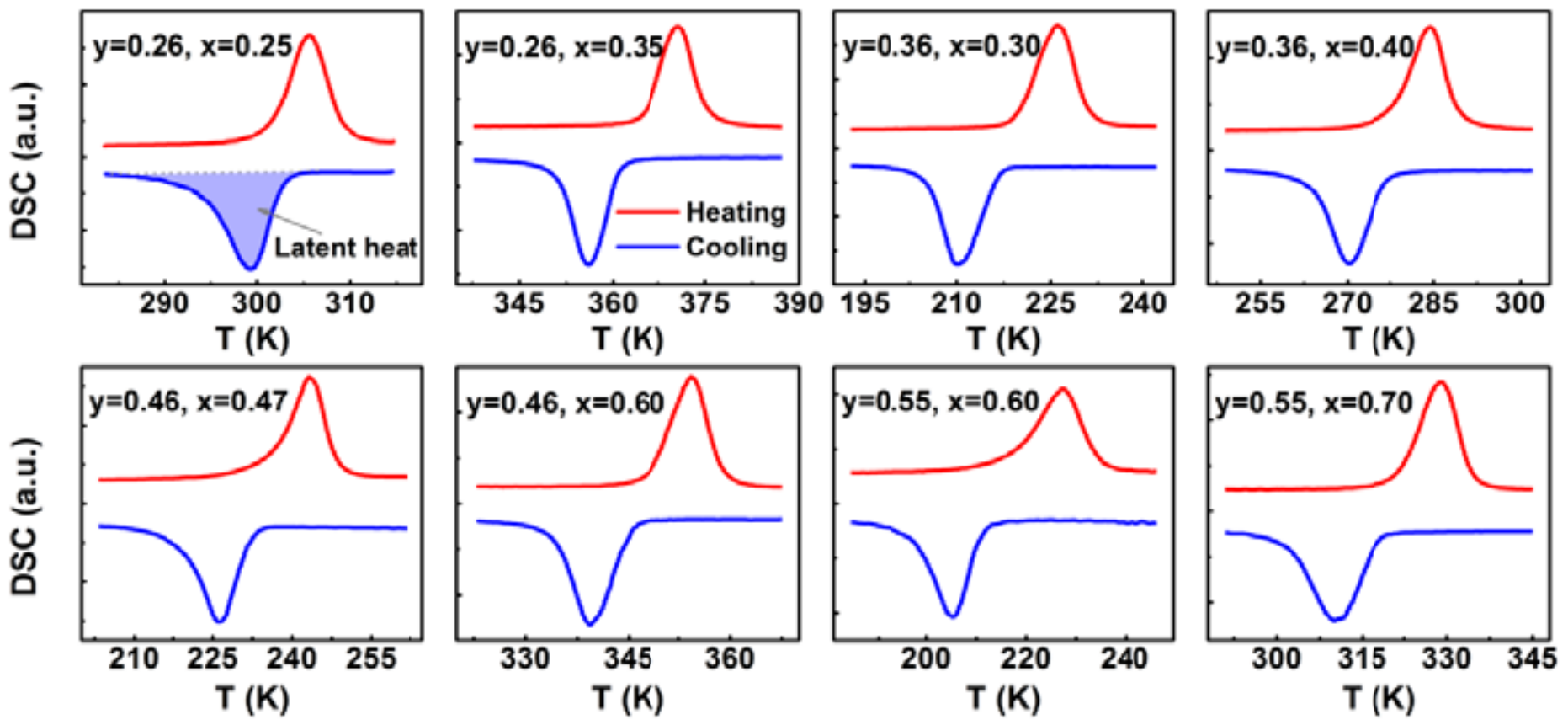

**Figure S6. Low-temperature DSC analysis for selected samples of $Mn_{1-y}Fe_yNiGe_{1-x}Si_x$ for (a) $y = 0.26$, (b) $y = 0.36$, (c) $y = 0.46$ and (d) $y = 0.55$.**

**Table S2. Parameters of $Mn_{1-y}Fe_yNiGe_{1-x}Si_x$. ΔM is magnetization change across the martensitic transitions, $T_t$ is martensitic transition temperature, $T_{cr}^{dn}$ and $T_{cr}^{up}$ are lower and upper critical temperatures of CTWs, respectively.**

| | $x$ | ΔM [emu g$^{-1}$] | $T_t$ [K] | $T_{cr}^{dn}$ [K] | $T_{cr}^{up}$ [K] | CTW width [K] |
|---|---|---|---|---|---|---|
| | **0.00** | - | **74** | **74** | | |
| | 0.05 | 58.1 | 113 | | | |
| | 0.15 | 66.4 | 218 | | | |
| **$y = 0.26$** | 0.25 | 62.2 | 299 | | | 360 |
| | 0.35 | 61.2 | 358 | | | |
| | 0.40 | 43.2 | 396 | | | |
| | **0.45** | - | **434** | | **434** | |
| | **0.15** | - | **51** | **51** | | |
| | 0.20 | 58.5 | 120 | | | |
| | 0.30 | 64.9 | 212 | | | |
| **$y = 0.36$** | 0.40 | 59.8 | 270 | | | 397 |
| | 0.50 | 54.5 | 348 | | | |
| | 0.60 | 44.9 | 426 | | | |
| | **0.63*** | - | **448*** | | **448*** | |
| | **0.30** | - | **40** | | | |
| | 0.35 | 57.4 | 118 | | | |
| | 0.40 | 65.1 | 161 | | | |
| **$y = 0.46$** | 0.47 | 61.9 | 230 | | | 405 |
| | 0.53 | 62.7 | 276 | | | |
| | 0.60 | 55.0 | 342 | | | |
| | 0.70 | 44.7 | 398 | | | |
| | **0.75** | 32.8 | **445** | | | |
| | **0.45** | - | **38** | **38** | | |
| | 0.50 | 53.1 | 114 | | | |
| ***y* = 0.55** | 0.60 | 59.5 | 209 | | | 386 |
| | 0.70 | 55.4 | 310 | | | |
| | 0.80 | 43.3 | 397 | | | |
| | **0.82*** | - | **424*** | | **424*** | |

*Extrapolated data.

**Table S3. Calculated magnetic moments of $Mn_{0.5}Fe_{0.5}NiGe_{0.5}Si_{0.5}$.***

| $M_{total}$ [$\mu_B$] | $M_{Mn}$ [$\mu_B$] | $M_{Fe}$ [$\mu_B$] | $M_{Ni}$ [$\mu_B$] | $M_{Ge}$ [$\mu_B$] | $M_{Si}$ [$\mu_B$] |
|---|---|---|---|---|---|
| **2.36** | 2.74 | 1.74 | 0.12 | -0.02 | -0.02 |
| | 2.76 | 1.76 | 0.14 | | |
| | | | 0.12 | | |
| | | | 0.12 | | |

*The energy favored atom configuration was chosen here. $\mu_B$ is Bohr magneton.

**Table S4. Latent heats (ΔE), corresponding magnetostructural transition temperatures ($T_t$) and estimated total entropy changes (ΔS) of several representative compositions of $Mn_{1-y}Fe_yNiGe_{1-x}Si_x$ deprived from DSC data shown Figure S5 and S6.**

| | ***x*** | **ΔE [J g$^{-1}$]** | **$T_t$ [K]** | **ΔS=ΔE/$T_t$ [J Kg$^{-1}$ K$^{-1}$]** |
|---|---|---|---|---|
| ***y* = 0.26** | 0.15 | 10.8 | 218 | 49.54 |
| | 0.25 | 16.2 | 299 | 54.18 |
| | 0.35 | 18.3 | 358 | 51.12 |
| | 0.60 | 16.4 | 561 | 29.23 |
| | 0.70 | 18.5 | 632 | 29.27 |
| | 0.85 | 18.1 | 861 | 21.02 |
| | 1.00 | 14.9 | 995 | 14.97 |
| ***y* = 0.36** | 0.30 | 12.9 | 212 | 60.85 |
| | 0.40 | 15.6 | 270 | 57.78 |
| | 0.60 | 26.6 | 426 | 62.44 |
| | 0.65 | 20.7 | 466 | 44.42 |
| | 0.70 | 20.3 | 494 | 41.09 |
| | 0.85 | 23.2 | 632 | 36.71 |
| | 1.00 | 19.3 | 866 | 22.29 |
| ***y* = 0.46** | 0.47 | 13.3 | 230 | 57.83 |
| | 0.60 | 19.0 | 342 | 55.56 |
| | 0.75 | 21.3 | 444 | 47.97 |
| | 0.80 | 21.8 | 488 | 44.67 |
| | 0.90 | 23.2 | 570 | 40.70 |
| | 1.00 | 29.7 | 648 | 45.83 |
| ***y* = 0.55** | 0.60 | 10.2 | 209 | 48.80 |
| | 0.70 | 19.7 | 310 | 63.55 |
| | 0.84 | 21.9 | 443 | 49.44 |
| | 0.90 | 22.1 | 487 | 45.38 |
| | 1.00 | 22.7 | 554 | 40.98 |

**Table S5. Absolute value of maximum entropy changes (|ΔS|) versus peak temperature for different magnetocaloric materials at field change ΔH = 20 kOe and ΔH = 50 kOe.**

| **$La(Fe,Si)_{13}$ based** | **T [K]** | **$\|\Delta S\|$ [J $Kg^{-1}$ $K^{-1}$] (ΔH=20 kOe)** | **T [K]** | **$\|\Delta S\|$ [J $Kg^{-1}$ $K^{-1}$] (ΔH=50 kOe)** | **Ref** |
|---|---|---|---|---|---|
| $LaFe_{11.4}Si_{1.6}$ | 208 | 14.3 | 208 | 19.4 | [1] |
| $LaFe_{11.83}Si_{1.17}$ | 175 | 21.2 | 175 | 27.8 | [2] |
| $LaFe_{11.7}Si_{1.3}$ | 183 | 22.9 | 183 | 26.0 | [2] |
| $LaFe_{11.6}Si_{1.4}$ | 188 | 22.9 | 188 | 26 | [2] |
| $LaFe_{11.5}Si_{1.5}$ | 194 | 20.8 | 194 | 24.8 | [2] |
| $LaFe_{11.4}Si_{1.6}$ | 199 | 14.2 | 199 | 18.7 | [2] |
| $LaFe_{11.3}Si_{1.7}$ | 206 | 11.9 | 206 | 17.6 | [2] |
| $LaFe_{11.2}Si_{1.8}$ | 210 | 7.5 | 210 | 13 | [2] |
| $LaFe_{11.0}Si_{2.0}$ | 221 | 4 | 221 | 7.9 | [2] |
| $La_{0.9}Pr_{0.1}Fe_{11.5}Si_{1.5}$ | 191 | 24 | 191 | 26.1 | [3] |
| $La_{0.7}Pr_{0.3}Fe_{11.5}Si_{1.5}$ | 185 | 28 | 185 | 30.5 | [3] |
| $La_{0.5}Pr_{0.5}Fe_{11.5}Si_{1.5}$ | 181 | 30 | 181 | 32.4 | [3] |
| $La_{0.9}Nd_{0.1}Fe_{11.5}Si_{1.5}$ | 192 | 23 | 192 | 25.9 | [3] |
| $La_{0.7}Nd_{0.3}Fe_{11.5}Si_{1.5}$ | 188 | 29 | 188 | 32 | [3] |
| $La_{0.7}Ce_{0.3}Fe_{11.5}Si_{1.5}$ | | | 173 | 23.8 | [3] |
| $La_{0.7}Pr_{0.3}Fe_{11.2}Si_{1.8}$ | 204 | 14.4 | 204 | 19.4 | [4] |
| $La_{0.7}Pr_{0.3}Fe_{11.0}Si_{2.0}$ | 218 | 6.2 | 218 | 11.4 | [4] |
| $La_{0.7}Pr_{0.3}Fe_{11.4}Si_{1.6}$ | 198 | 12 | 204 | 17 | [5] |
| $La_{0.7}Pr_{0.3}Fe_{11.34}Cu_{0.06}Si_{1.6}$ | | | 210 | 12 | [5] |
| $La_{0.7}Pr_{0.3}Fe_{11.06}Cu_{0.34}Si_{1.6}$ | | | 230 | 5 | [5] |
| $La_{0.8}Gd_{0.2}Fe_{11.4}Si_{1.6}$ | | | 200 | 14.8 | [6] |
| $LaFe_{10.7}Co_{0.8}Si_{1.5}$ | 285 | 7 | 285 | 13.5 | [7] |
| $La_{0.8}Pr_{0.2}Fe_{10.7}Co_{0.8}Si_{1.5}$ | 280 | 7.2 | 280 | 13.6 | [7] |
| $La_{0.6}Pr_{0.4}Fe_{10.7}Co_{0.8}Si_{1.5}$ | 274 | 7.4 | 274 | 14.2 | [7] |
| $La_{0.8}Pr_{0.5}Fe_{10.5}Co_{0.8}Si_{1.5}$ | 272 | 8.1 | 272 | 14.6 | [7] |
| $La(Fe_{0.96}Co_{0.04})_{1.9}Si_{1.1}$ | 243 | 16.4 | 243 | 23 | [8] |
| $La(Fe_{0.94}Co_{0.06})_{1.9}Si_{1.1}$ | 274 | 12.2 | 274 | 19.7 | [8] |
| $La(Fe_{0.92}Co_{0.08})_{1.9}Si_{1.1}$ | 301 | 8.7 | 301 | 15.6 | [8] |
| $LaFe_{11.2}Co_{0.7}Si_{1.1}$ | | | 274 | 20.3 | [9] |
| $LaFe_{10.98}Co_{0.22}Si_{1.8}$ | 242 | 6.3 | 242 | 11.5 | [10] |
| $LaFe_{11.12}Co_{0.71}Al_{1.17}$ | 279 | 4.6 | 279 | 9.1 | [10] |
| $La(Fe_{0.98}Co_{0.02})_{11.7}Al_{1.3}$ | 198 | 5.9 | 198 | 10.6 | [10] |
| $LaFe_{11.5}Si_{1.5}H_{0.3}$ | | | 224 | 17.4 | [11] |
| $LaFe_{11.5}Si_{1.5}H_{0.9}$ | | | 272 | 16.9 | [11] |
| $LaFe_{11.5}Si_{1.5}H_{1.3}$ | | | 288 | 17.0 | [11] |
| $LaFe_{11.5}Si_{1.5}H_{1.8}$ | | | 341 | 20.5 | [11] |
| $La(Fe_{0.99}Mn_{0.01})_{11.7}Si_{1.3}H_{\delta}$ | 336 | 16 | 336 | 23.4 | [12] |
| $La(Fe_{0.98}Mn_{0.02})_{11.7}Si_{1.3}H_{\delta}$ | 312 | 13.0 | 312 | 17.7 | [12] |
| $La(Fe_{0.97}Mn_{0.03})_{11.7}Si_{1.3}H_{\delta}$ | 287 | 11.0 | 287 | 15.9 | [12] |
| $LaFe_{11.5}Si_{1.5}C_{0.2}$ | 225 | 18 | 225 | 22.8 | [13] |
| $LaFe_{11.5}Si_{1.5}C_{0.5}$ | 241 | 7.4 | 241 | 12.7 | [13] |
| $La_{0.8}Gd_{0.2}Fe_{11.4}Si_{1.6}B_{0.03}$ | | | 204 | 16 | [6] |
| $La_{0.8}Gd_{0.2}Fe_{11.4}Si_{1.6}B_{0.06}$ | | | 205 | 15 | [6] |
| $La_{0.8}Gd_{0.2}Fe_{11.4}Si_{1.6}B_{0.3}$ | | | 222 | 6.6 | [6] |
| **Gd** | | | 295 | 10.5 | [14] |
| **$Gd_5(Si,Ge)_4$** | **T [K[** | **\|ΔS\| [J Kg$^{-1}$ K$^{-1}$[ (ΔH=20 kOe)** | **T [K]** | **\|ΔS\| [J Kg$^{-1}$ K$^{-1}$[ (ΔH=50 kOe)** | **Ref** |
| $Gd_5Si_2Ge_2$ | 276 | 14 | 276 | 18 | [15] |
| $Gd_5Si_{1.8}Ge_{2.2}$ | 243 | 13 | 243 | 20 | [16] |
| $Gd_5Si_4$ | 346 | 4.2 | 346 | 8.2 | [17] |
| $Gd_5Si_4$ | | | 336 | 8.2* | [18] |
| $Gd_5Si_{3.5}Ge_{0.5}$ | | | 331 | 7.3* | [18] |
| $Gd_5Si_3Ge$ | | | 323 | 8.7* | [18] |
| $Gd_5Si_{2.5}Ge_{1.5}$ | | | 313 | 9.4* | [18] |
| $Gd_5Si_{2.06}Ge_{1.94}$ | | | 306 | 9.4* | [18] |
| $Gd_5Si_2Ge_2$ | | | 276 | 19* | [18] |
| $Gd_5Si_{1.72}Ge_{2.28}$ | | | 246 | 40* | [18] |
| $Gd_5Si_1Ge_3$ | | | 140 | 72* | [18] |
| $Gd_5Si_{0.9}Ge_{3.1}$ | | | 130 | 32* | [18] |
| $Gd_5Si_{0.8}Ge_{3.2}$ | | | 121 | 22* | [18] |
| $Gd_5Si_{0.33}Ge_{3.67}$ | | | 68 | 38* | [18] |
| $Gd_5Si_{0.15}Ge_{3.85}$ | | | 40 | 24* | [18] |
| $Gd_5(Ge_{0.9175}Si_{0.0825})_4$ | | | 76 | 61 | [19] |
| $Gd_5(Ge_{0.75}Si_{0.25})_4$ | | | 143 | 69 | [19] |
| $Gd_5(Ge_{0.57}Si_{0.43})_4$ | | | 244 | 40 | [19] |
| $Gd_5(Ge_{0.5}Si_{0.5})_4$ | | | 277 | 19 | [19] |
| $Gd_5Si_{2.05}Ge_{1.95}$ | | | 279 | 9.2 | [20] |
| $Gd_5Ge_4$ | | | 20 | 17 | [18] |
| $Gd_5Ge_4$ | | | 40 | 26 | [19] |
| $Gd_{2.5}Tb_{2.5}Si_4$ | 280 | 4 | 280 | 8 | [17] |
| $Gd_3Tb_2Si_4$ | 296 | 4.1 | 296 | 8. | [17] |
| $Gd_{3.5}Tb_{1.5}Si_4$ | 311 | 4.5 | 311 | 9.5 | [17] |
| $Gd_4Tb_1Si_4$ | 318 | 4 | 318 | 8.2 | [17] |
| $Gd_{4.5}Tb_{0.5}Si_4$ | 334 | 4.1 | 334 | 8.2 | [17] |
| $Gd_4Tb_1Si_4$ | 291 | 3 | 291 | 6.1 | [17] |
| $Gd_{4.25}Tb_{0.75}Si_4$ | 304 | 3.6 | 304 | 7.5 | [17] |
| $Gd_{4.5}Tb_{0.5}Si_4$ | 316 | 4 | 316 | 8 | [17] |
| $Gd_5Ge_{1.9}Si_2Fe_{0.1}$ | | | 300 | 8 | [21] |
| $Gd_5Si_{2.02}Ge_{1.92}Mn_{0.06}$ | | | 275 | 11.6 | [20] |
| $Gd_5Si_{1.97}Ge_{1.87}Mn_{0.16}$ | | | 294 | 7 | [20] |
| $Gd_5Si_{1.985}Ge_{1.985}Ga_{0.03}$ | | | 276 | 8.4 | [22] |
| $Gd_5Si_{1.975}Ge_{1.975}Ga_{0.05}$ | | | 297 | 6.5 | [22] |
| $Gd_5Si_{1.935}Ge_{1.935}Ga_{0.13}$ | | | 297 | 6.1 | [22] |
| **$RE_5(Si,Ge)_4$** | **T [K]** | **$\|\Delta S\|$ [J $Kg^{-1}$ $K^{-1}$] ($\Delta H$=20 kOe)** | **T [K]** | **$\|\Delta S\|$ [J $Kg^{-1}$ $K^{-1}$] ($\Delta H$=50 kOe)** | **Ref** |
| $Tb_5Si_2Ge_2$ | 102 | 10 | 102 | 22 | [23] |
| $Tb_5Si_4$ | 225 | 5 | 225 | 10 | [23] |
| $Tb_5Ge_4$ | 90 | 1 | 90 | 4 | [23] |
| $Tb_5Si_2Ge_2$ | | | 102 | 13 | [24] |
| $Dy_5Si_4$ | | | 141 | 12.8 | [25] |
| $Dy_5(Si_{3.5}Ge_{0.5})$ | | | 136 | 11.9 | [25] |
| $Dy_5(Si_3Ge)$ | | | 65 | 33.2 | [25] |
| $Dy_5(Si_{2.3}Ge_{1.5})$ | | | 56 | 7.0 | [25] |
| $Dy_5(Si_2Ge_2)$ | | | 54 | 7.0 | [25] |
| $Dy_5(SiGe_3)$ | | | 50 | 7.0 | [25] |
| $Dy_5Ge_4$ | | | 46 | 7 | [25] |
| $Ho_5Si_2Ge_2$ | | | 25 | 58.5 | [26] |
| $Tb_5Si_4$ | 223 | 4.5 | 223 | 9 | [17] |
| **Hex RE compounds** | **T [K]** | **$\|\Delta S\|$ [J $Kg^{-1}$ $K^{-1}$] ($\Delta H$=20 kOe)** | **T [K]** | **$\|\Delta S\|$ [J $Kg^{-1}$ $K^{-1}$] ($\Delta H$=50 kOe)** | **Ref** |
| GdRhSn | 16 | 4 | 16 | 6.5 | [27] |
| HoPdAl (hexagonal) | 10 | 3 | 10 | 13.7 | [28] |
| HoPdAl (orthorhombic) | 12 | 11.4 | 12 | 22.8 | [28] |
| TbPdAl | 43 | 5.9 | 43 | 11.4 | [29] |
| HoCuSi | 8 | 16 | 8 | 34 | [30] |
| HoNiCuIn | 10 | 12.5 | 10 | 20.2 | [31] |
| $Er_2In$ | 46 | 7.9 | 46 | 16 | [32] |
| $Tb_2In$ | 165 | 3.5 | 165 | 6.6 | [33] |
| $Dy_2In$ | 130 | 4.6 | 130 | 9.2 | [34] |
| $Ho_2In$ | 85 | 5.0 | 85 | 11.2 | [35] |
| GdCoAl | 100 | 4.9 | 100 | 10.4 | [36] |
| TbCoAl | 70 | 5.3 | 70 | 10.5 | [36] |
| DyCoAl | 37 | 9.2 | 37 | 16.3 | [36] |
| HoCoAl | 10 | 12.5 | 10 | 21.5 | [36] |
| $(Gd_{0.5}Dy_{0.25}Er_{0.25})CoAl$ | 45 | 6.3 | 45 | 14 | [36] |
| **$RECo_2$** | **T [K]** | **$\|\Delta S\|$ [J $Kg^{-1}$ $K^{-1}$] ($\Delta H$=20 kOe)** | **T [K]** | **$\|\Delta S\|$ [J $Kg^{-1}$ $K^{-1}$] ($\Delta H$=50 kOe)** | **Ref** |
| $GdCo_{0.7}Mn_{1.3}$ | | | 300 | 3.35 | [37] |
| $GdCo_{0.6}Mn_{1.4}$ | | | 270 | 3.82 | [37] |
| $GdCo_{0.4}Mn_{1.6}$ | | | 210 | 3.86 | [37] |
| $GdCo_{0.2}Mn_{1.8}$ | | | 140 | 4.11 | [37] |
| $HoCo_2$ | | | 75 | 22.5 | [38] |
| $DyCo_2$ | | | 140 | 11 | [38] |
| $TbCo_2$ | | | 227 | 6.5 | [38] |
| $ErCo_2$ | | | 33.6 | 38 | [39] |
| $Er(Co_{0.95}Si_{0.05})_2$ | | | 60 | 22 | [39] |
| $Er(Co_{0.85}Si_{0.15})_2$ | | | 63 | 8 | [39] |
| $Er(Co_{0.975}Si_{0.025})_2$ | | | 48 | 32 | [38] |
| $Dy_{0.7}Y_{0.3}Co_2$ | | | 115 | 15 | [38] |
| $Dy_{0.9}Y_{0.1}Co_2$ | | | 130 | 14 | [38] |
| $Ho(Co_{0.95}Si_{0.05})_2$ | | | 100 | 20 | [38] |
| $Ho(Co_{0.975}Si_{0.025})_2$ | | | 80 | 23 | [38] |
| $ErCo_2$ | | | 38 | 23 | [40] |
| $Er(Co_{0.95}Fe_{0.05})_2$ | | | 70 | 13 | [40] |
| $Er(Co_{0.9}Fe_{0.1})_2$ | | | 150 | 6.5 | [40] |
| $ErCo_{1.8}Mn_{0.2}$ | | | 86 | 6.97 | [41] |
| $HoCo_{1.8}Mn_{0.2}$ | | | 156 | 5.36 | [41] |
| $DyCo_{1.8}Mn_{0.2}$ | | | 220 | 4.57 | [41] |
| $TbCo_{1.8}Mn_{0.2}$ | | | 312 | 3.2 | [41] |
| **MnFeX ($Fe_2P$ type)** | **T [K]** | **\|ΔS\| [J Kg$^{-1}$ K$^{-1}$] (ΔH=20 kOe)** | **T [K]** | **\|ΔS\| [J Kg$^{-1}$ K$^{-1}$] (ΔH=50 kOe)** | **Ref** |
| $MnFeP_{0.45}As_{0.55}$ | 305 | 14.5 | 309 | 18 | [42] |
| $MnFeP_{0.5}As_{0.5}$ | 288 | 16 | 288 | 21 | [43] |
| $Mn_{1.1}Fe_{0.9}P_{0.5}As_{0.5}$ | 280 | 26 | 280 | 28 | [43] |
| $Mn_{1.3}Fe_{0.7}P_{0.6}Si_{0.2}Ge_{0.15}$ | 246 | 2.5 | 246 | 5.7 | [44] |
| $Mn_{1.2}Fe_{0.8}P_{0.6}Si_{0.2}Ge_{0.15}$ | 285 | 5.3 | 285 | 11 | [44] |
| $Mn_{1.1}Fe_{0.9}P_{0.6}Si_{0.2}Ge_{0.15}$ | 325 | 12 | 325 | 16 | [44] |
| Mn $FeP_{0.6}Si_{0.2}Ge_{0.15}$ | 333 | 11 | 333 | 17.5 | [44] |
| $Mn_{1.2}Fe_{0.75}P_{0.5}Si_{0.5}$ | | | 304 | 31 | [45] |
| $Mn_{1.25}Fe_{0.70}P_{0.5}Si_{0.5}$ | | | 285 | 27 | [45] |
| $Mn_{1.3}Fe_{0.65}P_{0.5}Si_{0.5}$ | | | 269 | 21 | [45] |
| $Mn_{1.25}Fe_{0.7}P_{0.55}Si_{0.45}$ | | | 246 | 33 | [45] |
| $Mn_{1.25}Fe_{0.7}P_{0.45}Si_{0.55}$ | | | 323 | 19 | [45] |
| $Mn_{1.2}Fe_{0.8}P_{0.75}Ge_{0.25}$ | 288 | 20.3 | 288 | 25 | [46] |
| **MnAs** | **T [K]** | **\|ΔS\| [J Kg$^{-1}$ K$^{-1}$] (ΔH=20 kOe)** | **T [K]** | **\|ΔS\| [J Kg$^{-1}$ K$^{-1}$] (ΔH=50 kOe)** | **Ref** |
| MnAs | 315 | 38 | 315 | 41 | [47] |
| MnAs | | | 317 | 41 | [48] |
| $MnAs_{0.95}Sb_{0.05}$ | | | 308 | 32 | [48] |
| $MnAs_{0.85}Sb_{0.15}$ | | | 260 | 30 | [48] |
| $MnAs_{0.70}Sb_{0.30}$ | 225 | 23 | 225 | 27 | [48] |
| $MnAs_{0.6}Sb_{0.4}$ | 214 | 7 | 214 | 14 | [48] |
| $MnAs_{0.95}Sb_{0.05}$ | 310 | 27 | 310 | 33 | [47] |
| $MnAs_{0.9}Sb_{0.1}$ | 285 | 25 | 285 | 32 | [47] |
| $Mn_{0.994}Cr_{0.006}As$ | | | 292 | 13.7 | [49] |
| $Mn_{0.99}Cr_{0.01}As$ | | | 267 | 20.2 | [49] |

| MM’X ($Ni_2In$ type) | T [K] | \|ΔS\| [J $Kg^{-1}$ $K^{-1}$] (ΔH=20 kOe) | T [K] | \|ΔS\| [J $Kg^{-1}$ $K^{-1}$] (ΔH=50 kOe) | Ref |
|---|---|---|---|---|---|
| $MnNi_{0.77}Fe_{0.23}Ge$ | 265 | 7 | 265 | 19 | [50] |
| $Mn_{0.82}Fe_{0.18}NiGe$ | 203 | 13 | 203 | 31 | [50] |
| $Mn_{0.93}Cr_{0.03}CoGe$ | | | 300 | 22 | [51] |
| $Mn_{0.94}Ti_{0.06}CoGe$ | | | 232 | 11.3 | [52] |
| $MnCoGe_{0.99}Al_{0.01}$ | | | 357 | 29.2 | [53] |
| $MnCoGe_{0.98}Al_{0.02}$ | | | 323 | 35.9 | [53] |
| $MnCoGe_{0.97}Al_{0.03}$ | | | 180 | 5.7 | [53] |
| $Mn_{0.9}Co_{0.1}NiGe$ | | | 236 | 40 | [54] |
| $Mn_{0.965}CoGe$ | 289 | 10 | 289 | 26 | [55] |
| $Mn_{1.05}Ni_{0.85}Ge$ | 135 | 14 | 135 | 27 | [56] |
| $Mn_{1.045}Ni_{0.855}Ge$ | 160 | 12 | 160 | 25 | [56] |
| $(MnNiSi)_{0.56}(FeNiGe)_{0.44}$ | | | 187 | 11.5 | [57] |
| $(MnNiSi)_{0.54}(FeNiGe)_{0.46}$ | | | 192 | 10.8 | [57] |
| $(MnNiSi)_{0.52}(FeNiGe)_{0.48}$ | | | 174 | 8.4 | [57] |
| $Mn_{0.89}Cr_{0.11}NiGe$ | 275 | 11 | 275 | 28 | [58] |
| $Mn_{0.92}Cu_{0.08}CoGe$ | 321 | 21 | 321 | 53.3 | [59] |
| $Mn_{0.91}Cu_{0.09}CoGe$ | 289 | 16 | 289 | 41.2 | [59] |
| $Mn_{0.9}Cu_{0.1}CoGe$ | 249 | 15 | 249 | 36.4 | [59] |
| $MnCoGeB_{0.02}$ | 287 | 21 | 287 | 47.3 | [60] |
| $MnCoGeB_{0.03}$ | 275 | 16 | 275 | 37.7 | [60] |
| $MnCoGeB_{0.05}$ | | | 260 | 3.4 | [60] |
| **Heusler alloys** | **T [K]** | **\|ΔS\| [J $Kg^{-1}$ $K^{-1}$] (ΔH=20 kOe)** | **T [K]** | **\|ΔS\| [J $Kg^{-1}$ $K^{-1}$] (ΔH=50 kOe)** | **Ref** |
| $Ni_{50}Mn_{37}Sn_{13}$ | | | 189 | 18 | [61] |
| $Ni_{50}Mn_{35}Sn15$ | | | 307 | 15 | [61] |
| $Ni_{52.6}Mn_{23.1}Ga_{24.3}$ | 304 | 6.5 | 304 | 18 | [62] |
| $Ni_2Mn_{0.755}Cu_{0.245}Ga$ | 298 | 29 | 298 | 45 | [63] |
| $Ni_2Mn_{0.75}Cu_{0.25}Ga$ | 308 | 29 | 308 | 65 | [63] |
| $Ni_2Mn_{0.74}Cu_{0.26}Ga$ | 302 | 25 | 302 | 43 | [63] |
| $Ni_{50}Mn_{34}In_{16}$ | 243 | 3 | 243 | 8 | [64] |
| $Ni_{50}Mn_{34}In_{14}Ga_2$ | 275 | 3 | 275 | 9 | [64] |
| $Ni_2Mn_{1.2}Ga_{0.8}$ | | | 352 | 7.3 | [65] |
| $Ni_{1.9}Mn_{1.3}Ga_{0.8}$ | | | 365 | 9.6 | [65] |
| $Ni_{43}Mn_{43}Co_3Sn_{11}$ | 188 | 18 | 188 | 33 | [66] |
| $Ni_{50}Mn_{37}Sn_{13}$ | | | 280 | 0.2 | [67] |
| $Ni_{49}M_{n38}Sn_{13}$ | | | 270 | 2.2 | [67] |
| $Ni_{48}Mn_{39}Sn_{13}$ | | | 226 | 12 | [67] |
| $Ni_{47}Mn_{40}Sn_{13}$ | | | 187 | 34 | [67] |
| $Ni_{46}Mn_{41}Sn_{13}$ | | | 132 | 11 | [67] |
| $Ni_{43}Mn_{46}Sn_{11}$ | | | 205 | 10.4 | [68] |
| $Ni_{43}Mn_{46}Sn_{10.5}Al_{0.5}$ | | | 236 | 8.1 | [68] |
| $Ni_{43}Mn_{46}Sn_{10}Al_1$ | | | 264 | 3.9 | [68] |
| $Ni_{43}Mn_{46}Sn_9Al_2$ | | | 297 | 1.4 | [68] |
| $Ni_{51}Mn_{32.8}In_{16.2}$ | | | 253 | 19.2 | [69] |
| $Ni_{51}Mn_{32.8}In_{16.2}H_{1.4}$ | | | 245 | 17.5 | [69] |
| $Ni_{44.4}Mn_{44.1}Sn_{11.5}$ | 260 | 10.6 | | | [70] |
| $Ni_{44.2}Mn_{39.3}Fe_{4.9}Sn_{11.6}$ | 284 | 6.4 | | | [70] |
| $Ni_{44.5}Mn_{37}Fe_{6.7}Sn_{11.8}$ | 293 | 5.6 | | | [70] |
| $Ni_{49}Co_1Mn_{37}Sn_{13}$ | 290 | 3 | 290 | 9 | [71] |
| $Ni_{47}Co_3Mn_{37}Sn_{13}$ | 255 | 3 | 255 | 9 | [71] |
| $Ni_{49}Fe_1Mn_{37}Sn_{13}$ | 240 | 6 | 240 | 16 | [71] |
| $Ni_{47}Fe_3Mn_{37}Sn_{13}$ | 175 | 11 | 175 | 30 | [71] |
| $Ni_{55.4}Mn_{20}Ga_{24.6}$ | 313 | 40 | 313 | 85 | [72] |
| $Ni_{51}Mn_{32.4}In_{16.6}$ | 246 | 16 | 246 | 17.2 | [69] |
| $Ni_{51}Mn_{32.4}In_{16.6}H_{5.2}$ | 219 | 9 | 219 | 13 | [69] |
| $Ni_{43}Mn_{42}Co_4Sn_{11}$ | | | 220 | 19 | [73] |
| $Ni_{42}Co_8Mn_{30}Fe_2Ga_{18}$ | 205 | 11 | 205 | 31 | [74] |
| $Ni_{46}Co_4Mn_{38}Sb_{12}$ | | | 295 | 32.3 | [75] |
| $Ni_{52}Mn_{26}Ga_{22}$ | 355 | 16 | 355 | 30 | [76] |
| **Present work** | **T [K]** | **\|ΔS\| [J $Kg^{-1}$ $K^{-1}$] (ΔH=20 kOe)** | **T [K]** | **\|ΔS\| [J $Kg^{-1}$ $K^{-1}$] (ΔH=50 kOe)** | |
| $Mn_{0.64}Fe_{0.36}NiGe_{0.8}Si_{0.2}$ | 127 | 10.7 | 127 | 27.9 | |
| $Mn_{0.64}Fe_{0.36}NiGe_{0.7}Si_{0.3}$ | 219 | 11.4 | 219 | 33.6 | |
| $Mn_{0.64}Fe_{0.36}NiGe_{0.6}Si_{0.4}$ | 277 | 11.6 | 277 | 30.5 | |
| $Mn_{0.64}Fe_{0.36}NiGe_{0.5}Si_{0.5}$ | 353 | 9.6 | 353 | 24.7 | |
| $Mn_{0.64}Fe_{0.36}NiGe_{0.4}Si_{0.6}$ | 437 | 7.6 | 437 | 27.4 | |

*the data calculated from the entropy change per unit volume with assuming density of 7.5 g $cm^{-3}$

# References

[1] F.-X. Hu, B.-G. Shen, J.-R. Sun, Z.-H. Cheng, G.-H. Rao, X.-X. Zhang, *Appl. Phys. Lett.* **2001**, *78*, 3675.

[2] B. G. Shen, J. R. Sun, F. X. Hu, H. W. Zhang, Z. H. Cheng, *Adv. Mater.* **2009**, *21*, 4545.

[3] S. Jun, L. Yang-Xian, S. Ji-Rong, S. Bao-Gen, *Chin. Phys. B* **2009**, *18*, 2058.

[4] J. Shen, Y.-X. Li, Q.-Y. Dong, J.-R. Sun, *J. Magn. Magn. Mater.* **2009**, *321*, 2336.

[5] M. F. Md Din, J. L. Wang, R. Zeng, P. Shamba, J. C. Debnath, S. X. Dou, *Intermetallics* **2013**, *36*, 1.

[6] P. Shamba, R. Zeng, J. L. Wang, S. J. Campbell, S. X. Dou, *J. Magn. Magn. Mater.* **2013**, *331*, 102.

[7] J. Shen, B. Gao, Q.-Y. Dong, Y.-X. Li, F.-X. Hu, J.-R. Sun, B.-G. Shen, *J. Phys. D - Appl. Phys.* **2008**, *41*, 245005.

[8] F. X. Hu, J. Gao, X. L. Qian, M. Ilyn, A. M. Tishin, J. R. Sun, B. G. Shen, *J. Appl. Phys.* **2005**, *97*, 10M303.

[9] F.-X. Hu, B.-g. Shen, J.-r. Sun, G.-j. Wang, Z.-h. Cheng, *Appl. Phys. Lett.* **2002**, *80*, 826.

[10] F.-X. Hu, S. B.-g. Shen, J.-r. Sun, X.-X. Zhang, *Chin. Phys.* **2000**, *9*, 550.

[11] Y. F. Chen, F. Wang, B. G. Shen, F. X. Hu, J. R. Sun, G. J. Wang, Z. H. Cheng, *J. Phys. - Condens. Matter* **2003**, *15*, L161.

[12] W. Fang, C. Yuan-Fu, W. Guang-Jun, S. Ji-Rong, S. Bao-Gen, *Chin. Phys.* **2003**, *12*, 911.

[13] Y.-f. Chen, F. Wang, B.-G. Shen, J.-R. Sun, G.-J. Wang, F.-X. Hu, Z.-H. Cheng, T. Zhu, *J. Appl. Phys.* **2003**, *93*, 6981.

[14] V. K. Pecharsky, K. A. Gschneidner Jr, *J. Magn. Magn. Mater.* **1999**, *200*, 44.

[15] V. K. Pecharsky, J. K. A. Gschneidner, *Phys. Rev. Lett.* **1997**, *78*, 4494.

[16] F. Casanova, X. Batlle, A. Labarta, J. Marcos, L. Mañosa, A. Planes, *Phys. Rev. B* **2002**, *66*, 100401.

[17] Y. I. Spichkin, V. K. Pecharsky, K. A. Gschneidner, *J. Appl. Phys.* **2001**, *89*, 1738.

[18] K. A. Gschneidner, V. K. Pecharsky, *Annu. Rev. Mater. Sci.* **2000**, *30*, 387.

[19] V. K. Pecharsky, K. A. Gschneidner, *Appl. Phys. Lett.* **1997**, *70*, 3299.

[20] E. Yüzüak, I. Dincer, Y. Elerman, *J. Rare Earths* **2010**, *28*, 477.

[21] V. Provenzano, A. J. Shapiro, R. D. Shull, *Nature* **2004**, *429*, 853.

[22] S. Aksoy, A. Yucel, Y. Elerman, T. Krenke, M. Acet, X. Moya, L. Mañosa, *J. Alloys Compd.* **2008**, *460*, 94.

[23] L. Morellon, C. Magen, P. A. Algarabel, M. R. Ibarra, C. Ritter, *Appl. Phys. Lett.* **2001**, *79*, 1318.

[24] L. Morellon, Z. Arnold, C. Magen, C. Ritter, O. Prokhnenko, Y. Skorokhod, P. A. Algarabel, M. R. Ibarra, J. Kamarad, *Phys. Rev. Lett.* **2004**, *93*, 137201.

[25] V. V. Ivtchenko, V. K. Pecharsky, K. A. J. Gschneidner, in *Advanced Cryogenic Engineering*, Vol. 46, 2000, 405.

[26] N. P. Thuy, Y. Y. Chen, Y. D. Yao, C. R. Wang, S. H. Lin, J. C. Ho, T. P. Nguyen, P. D. Thang, J. C. P. Klaasse, N. T. Hien, L. T. Tai, *J. Magn. Magn. Mater.* **2003**, *262*, 432.

[27] S. Gupta, K. G. Suresh, A. K. Nigam, Y. Mudryk, D. Paudyal, V. K. Pecharsky, K. A. Gschneidner, *J. Alloys Compd.* **2014**, *613*, 280.

[28] Z. Xu, B. Shen, *Sci. China - Technol. Sci.* **2011**, *55*, 445.

[29] J. Shen, Z.-Y. Xu, H. Zhang, X.-Q. Zheng, J.-F. Wu, F.-X. Hu, J.-R. Sun, B.-g. Shen, *J. Magn. Magn. Mater.* **2011**, *323*, 2949.

[30] J. Chen, B. G. Shen, Q. Y. Dong, F. X. Hu, J. R. Sun, *Appl. Phys. Lett.* **2010**, *96*, 152501.

[31] Z.-J. Mo, J. Shen, L.-Q. Yan, C.-C. Tang, X.-N. He, X. Zheng, J.-F. Wu, J.-R. Sun, B.-G. Shen, *J. Magn. Magn. Mater.* **2014**, *354*, 49.

[32] H. Zhang, B. G. Shen, Z. Y. Xu, J. Chen, J. Shen, F. X. Hu, J. R. Sun, *J. Alloys Compd.* **2011**, *509*, 2602.

[33] Q. Zhang, J. H. Cho, J. Du, F. Yang, X. G. Liu, W. J. Feng, Y. J. Zhang, J. Li, Z. D. Zhang, *Solid State Commun.* **2009**, *149*, 396.

[34] Q. Zhang, X. G. Liu, F. Yang, W. J. Feng, X. G. Zhao, D. J. Kang, Z. D. Zhang, *J. Phys. D - Appl. Phys.* **2009**, *42*, 055011.
[35] Q. Zhang, J. H. Cho, B. Li, W. J. Hu, Z. D. Zhang, *Appl. Phys. Lett.* **2009**, *94*, 182501.
[36] X. X. Zhang, F. W. Wang, G. H. Wen, *J. Phys. - Condens. Matter* **2001**, *13*, L747.
[37] J. Y. Zhang, J. Luo, J. B. Li, J. K. Liang, Y. C. Wang, L. N. Ji, Y. H. Liu, G. H. Rao, *Solid State Commun.* **2007**, *143*, 541.
[38] N. H. Duc, D. T. Kim Anh, P. E. Brommer, *Phys. B* **2002**, *319*, 1.
[39] N. H. Duc, P.E. Brommer, K. H. J. Buschow, *Handbook on Magnetic Materials*, Vol. 17, 1999, Ch. 5.
[40] X. B. Liu, Z. Altounian, *J. of Appl. Phys.* **2008**, *103*, 07B304.
[41] A. K. Pathak, I. Dubenko, S. Stadler, N. Ali, *J. Magn. Magn. Mater.* **2011**, *323*, 2436.
[42] O. Tegus, E. Bruck, K. H. J. Buschow, F. R. de Boer, *Nature* **2002**, *415*, 150.
[43] E. Brück, O. Tegus, X. W. Li, F. R. de Boer, K. H. J. Buschow, *Phys. B* **2003**, *327*, 431.
[44] G. F. Wang, Z. R. Zhao, L. Song, O. Tegus, *J. Alloys Compd.* **2013**, *554*, 208.
[45] N. H. Dung, L. Zhang, Z. Q. Ou, E. Brück, *Scr. Mater.* **2012**, *67*, 975.
[46] N. T. Trung, J. C. P. Klaasse, O. Tegus, D. T. Cam Thanh, K. H. J. Buschow, E. Brück, *J. Phys. D - Appl. Phys.* **2010**, *43*, 015002.
[47] H. Wada, Y. Tanabe, *Appl. Phys. Lett.* **2001**, *79*, 3302.
[48] H. Wada, K. Taniguchi, Y. Tanabe, *Mater. Trans.* **2002**, *43*, 73.
[49] N. K. Sun, W. B. Cui, D. Li, D. Y. Geng, F. Yang, Z. D. Zhang, *Appl. Phys. Lett.* **2008**, *92*, 072504.
[50] E. Liu, W. Wang, L. Feng, W. Zhu, G. Li, J. Chen, H. Zhang, G. Wu, C. Jiang, H. Xu, F. de Boer, *Nat. Commun.* **2012**, *3*, 873.
[51] L. Caron, N. T. Trung, E. Brück, *Phys. Rev. B* **2011**, *84*, 020414.
[52] P. Shamba, J. L. Wang, J. C. Debnath, S. J. Kennedy, R. Zeng, M. F. Din, F. Hong, Z. X. Cheng, A. J. Studer, S. X. Dou, *J. Phys. - Condens. Matter* **2013**, *25*, 056001.
[53] L. F. Bao, F. X. Hu, R. R. Wu, J. Wang, L. Chen, J. R. Sun, B. G. Shen, L. Li, B. Zhang, X. X. Zhang, *J. Phys. D - Appl. Phys.* **2014**, *47*, 055003.
[54] E. K. Liu, H. G. Zhang, G. Z. Xu, X. M. Zhang, R. S. Ma, W. H. Wang, J. L. Chen, H. W. Zhang, G. H. Wu, L. Feng, X. X. Zhang, *Appl. Phys. Lett.* **2013**, *102*, 122405.
[55] E. K. Liu, W. Zhu, L. Feng, J. L. Chen, W. H. Wang, G. H. Wu, H. Y. Liu, F. B. Meng, H. Z. Luo, Y. X. Li, *Europhys. Lett.* **2010**, *91*, 17003.
[56] C. L. Zhang, D. H. Wang, Q. Q. Cao, Z. D. Han, H. C. Xuan, Y. W. Du, *Appl. Phys. Lett.* **2008**, *93*, 122505.
[57] C. L. Zhang, D. H. Wang, Z. D. Han, B. Qian, H. F. Shi, C. Zhu, J. Chen, T. Z. Wang, *Appl. Phys. Lett.* **2013**, *103*, 132411.
[58] A. P. Sivachenko, V. I. Mityuk, V. I. Kamenev, A. V. Golovchan, V. I. Val'kov, I. F. Gribanov, *Low Temp. Phys.* **2013**, *39*, 1051.
[59] T. Samanta, I. Dubenko, A. Quetz, S. Stadler, N. Ali, *Appl. Phys. Lett.* **2012**, *101*, 242405.
[60] N. T. Trung, L. Zhang, L. Caron, K. H. J. Buschow, E. Brück, *Appl. Phys. Lett.* **2010**, *96*, 172504.
[61] T. Krenke, E. Duman, M. Acet, E. F. Wassermann, X. Moya, L. Manosa, A. Planes, *Nat. Mater.* **2005**, *4*, 450.
[62] F.-X. Hu, B.-G. Shen, J.-R. Sun, G.-H. Wu, *Phys. Rev. B* **2001**, *64*, 132412.
[63] S. Stadler, M. Khan, J. Mitchell, N. Ali, A. M. Gomes, I. Dubenko, A. Y. Takeuchi, A. P. Guimarães, *Appl. Phys. Lett.* **2006**, *88*, 192511.
[64] S. Aksoy, T. Krenke, M. Acet, E. F. Wassermann, X. Moya, L. s. Mañosa, A. Planes, *Appl. Phys. Lett.* **2007**, *91*, 241916.
[65] F. Albertini, A. Paoluzi, L. Pareti, M. Solzi, L. Righi, E. Villa, S. Besseghini, F. Passaretti, *J. Appl. Phys.* **2006**,

*100*, 023908.
[66] B. Gao, F. X. Hu, J. Shen, J. Wang, J. R. Sun, B. G. Shen, *J. Magn. Magn. Mater.* **2009**, *321*, 2571.
[67] S. E. Muthu, N. V. R. Rao, M. M. Raja, D. M. R. Kumar, D. M. Radheep, S. Arumugam, *J. Phys. D - Appl. Phys.* **2010**, *43*, 425002.
[68] J. Chen, Z. Han, B. Qian, P. Zhang, D. Wang, Y. Du, *J. Magn. Magn. Mater.* **2011**, *323*, 248.
[69] F. X. Hu, J. Wang, L. Chen, J. L. Zhao, J. R. Sun, B. G. Shen, *Appl. Phys. Lett.* **2009**, *95*, 112503.
[70] J. L. Yan, Z. Z. Li, X. Chen, K. W. Zhou, S. X. Shen, H. B. Zhou, *J. Alloys Compd.* **2010**, *506*, 516.
[71] T. Krenke, E. P. Duman, M. Acet, X. Moya, L. S. Mañosa, A. Planes, *J. Appl. Phys.* **2007**, *102*, 033903.
[72] M. Pasquale, C. P. Sasso, L. H. Lewis, L. Giudici, T. Lograsso, D. Schlagel, *Phys. Rev. B* **2005**, *72*, 094435.
[73] N. M. Bruno, C. Yegin, I. Karaman, J.-H. Chen, J. H. Ross, J. Liu, J. Li, *Acta Mater.* **2014**, *74*, 66.
[74] A. K. Pathak, I. Dubenko, H. E. Karaca, S. Stadler, N. Ali, *Appl. Phys. Lett.* **2010**, *97*, 062505.
[75] A. K. Nayak, K. G. Suresh, A. K. Nigam, A. A. Coelho, S. Gama, *J. Appl. Phys.* **2009**, *106*, 053901.
[76] Z. Li, Y. Zhang, C. F. Sánchez-Valdés, J. L. Sánchez Llamazares, C. Esling, X. Zhao, L. Zuo, *Appl. Phys. Lett.* **2014**, *104*, 044101.